\documentclass[a4paper,11pt]{article}
\usepackage{jheppub}

\usepackage{lmodern}
\usepackage{amsmath,amssymb,amsfonts,mathtools,amsthm}
\usepackage{array,booktabs,multirow}
\usepackage{xcolor}
\usepackage{enumitem}
\usepackage{tikz}
\usetikzlibrary{fit,positioning}

\DeclareMathOperator{\tr}{tr}
\DeclareMathOperator{\Vol}{Vol}
\DeclareMathOperator*{\Res}{Res}
\newcommand{\ap}{\alpha'}
\newcommand{\dd}{\mathrm d}
\newcommand{\dlog}{\dd\log}
\newcommand{\R}{\mathbb R}
\newcommand{\Rpos}{\mathbb R_{>0}}
\newcommand{\Conv}{\operatorname{Conv}}
\newcommand{\Newt}{\operatorname{Newt}}
\newcommand{\bfe}{\mathbf{e}}
\newcommand{\calI}{\mathcal I}
\newcommand{\calC}{\mathcal C}
\newcommand{\calB}{\mathcal B}
\newcommand{\calJ}{\mathcal J}
\newcommand{\calS}{\mathcal S}

\newcommand{\zono}{\mathrm{zono}}
\newcommand{\Zono}{\mathrm{Zono}}
\newcommand{\cube}{\mathrm{cube}}
\newcommand{\Inc}{\operatorname{Inc}}

\newcommand{\corrG}{\langle G\rangle}

\newcommand{\Trop}{\operatorname{Trop}}

\newcommand{\calG}{\mathcal G}

\newcommand{\Asso}{\mathrm{\mathcal{A}sso}}
\newcommand{\Cube}{\mathrm{\mathcal{C}ube}}

\theoremstyle{plain}
\newtheorem{proposition}{Proposition}[section]
\theoremstyle{remark}

\title{Tropical and Stringy Integrals for In-In Correlators}

\author[a,b,c]{Song He,}
\author[a,c]{Xiang Li,}
\author[a,c]{Yong-Xiang Su}
\author[a,d]{and Fan Zhu}

\affiliation[a]{New Cornerstone Laboratory, Institute of Theoretical Physics, Chinese Academy of Sciences, Beijing 100190, China}
\affiliation[b]{School of Fundamental Physics and Mathematical Sciences, Hangzhou Institute for Advanced Study and ICTP-AP, UCAS, Hangzhou 310024, China}
\affiliation[c]{School of Physical Sciences, University of Chinese Academy of Sciences, No.19A Yuquan Road, Beijing 100049, China}
\affiliation[d]{Graduate School of China Academy of Engineering Physics, No. 10 Xibeiwang East Road, Haidian District, Beijing 100193, P.R. China}

\emailAdd{songhe@itp.ac.cn}
\emailAdd{lixiang@itp.ac.cn}
\emailAdd{anonym20031201@gmail.com}
\emailAdd{zhufan25@gscaep.ac.cn}

\abstract{
	We introduce tropical and stringy integrals for fixed-graph contributions to cosmological in-in correlators of conformally coupled scalars.  The full-time representation factorizes into a graph-dependent vertex-space Laplace integral, with one real variable for each graph vertex, and an elementary edge-space Laplace integral, with one real variable for each internal edge.  The vertex-space exponent is a sum of absolute values associated with sites and relative edge times; as a piecewise-linear function, it is the support function of the in-in zonotope.  Each absolute value is also the tropical limit of a positive Laurent binomial.  Retaining these binomials before tropicalization defines a finite-$\ap$ vertex-space stringy integral, so both the polytope and its stringy integral are read directly from the physical time integral.  In the $\ap{\to}0$ limit this integral becomes the normalized dual volume of the in-in zonotope, while the edge-space factor deforms independently into a product of beta integrals and restores the elementary propagator normalization.  We derive the field-theory rational form from augmented-graph chambers, as well as exact finite-$\ap$ parallel-edge reduction, factorization formulas for edge-energy and partial-energy poles, and even descendant towers. As an alternative geometric realization of fixed-graph correlators, we find an ambient Minkowski-sum and stringy-integral realization of the graph correlahedron for a tree graph as the so-called {\it graph cubeahedron} of its line graph. For completeness, we also record the logarithmic critical equations and generic reference degrees of the associated affine divisor arrangement.
}
\date{\today}

\begin{document}
\maketitle

\section{Introduction}

Recent work on scattering amplitudes has shown that locality and factorization can often be encoded directly by combinatorial geometry, rather than recovered only after summing Feynman diagrams.  Positive geometries and their canonical forms, together with stringy canonical forms and binary geometries, provide concrete realizations of this idea \cite{Arkani-Hamed:2017tmz,Arkani-Hamed:2017mur,ArkaniHamedHeSalvatoriThomas2019,ArkaniHamedHeLamThomas2019,ArkaniHamedHeLamThomas2023}.  This motivates asking whether cosmological observables admit a similarly direct tropical and stringy description.

The prototype for the relation between amplitudes, positive geometries, and stringy integrals is the color-ordered planar tree amplitude of $\operatorname{Tr}(\phi^3)$ theory.  For the cyclic ordering $(1,2,\ldots,n)$,
\begin{equation}
	m_n^{\operatorname{Tr}(\phi^3)} =\sum_{T\in\operatorname{Triang}(n)} \prod_{(i,j)\in T}\frac{1}{X_{ij}} =\Omega(\mathcal A_{n-3}) =\lim_{\ap\to0}\calI^{\operatorname{Tr}(\phi^3)}_n(\ap).
	\label{eq:intro-phi3-abhy-string}
\end{equation}
Here $T$ runs over triangulations of the $n$-gon and $(i,j)\in T$ over its internal diagonals.  The planar propagator variables are $X_{ij}{=}(k_i+k_{i+1}+\cdots+k_{j-1})^2$. The same rational function is the canonical function of the $(n{-}3)$-dimensional ABHY associahedron $\mathcal A_{n-3}$ and the field-theory limit of the finite-$\ap$ integral in \eqref{eq:intro-phi3-abhy-string} \cite{Arkani-Hamed:2017tmz,Arkani-Hamed:2017mur}. More recently, the surface-based curve-integral formalism extended this $\operatorname{Tr}(\phi^3)$ story to arbitrary loop order and all orders in the topological 't Hooft expansion.  Its tropical formulation expresses loop-integrated amplitudes as curve integrals built from headlight functions \cite{ArkaniHamedFrostSalvatoriPlamondonThomas2023Counting, ArkaniHamedFrostSalvatoriPlamondonThomas2023Multiplicity, ArkaniHamedFigueiredoFrostSalvatori2024}.

On the ordered component of $\mathcal M_{0,n}(\mathbb R)$ with ordering $(1,2,\ldots,n)$, choose the standard positive coordinates $z_2,\ldots,z_{n-2}>0$.  The finite-$\ap$ integral in \eqref{eq:intro-phi3-abhy-string} then takes the positive Laurent-polynomial form \cite{ArkaniHamedHeLamThomas2019}
\begin{equation}
	\begin{aligned}
		\calI^{\operatorname{Tr}(\phi^3)}_n(\ap) &:=(\ap)^{n-3} \int_{\Rpos^{n-3}}\prod_{a=2}^{n-2}\frac{\dd z_a}{z_a}\, \prod_{a=2}^{n-2}z_a^{\ap X_{a n}} \prod_{1\leq i<i+1<j\leq n-1} F_{ij}(\mathbf z)^{-\ap c_{ij}},\\[5pt]
		F_{ij}(\mathbf z) &:=1+z_{i+1}+z_{i+1}z_{i+2}+\cdots+z_{i+1}z_{i+2}\cdots z_{j-1}.
	\end{aligned}
	\label{eq:intro-z-integral}
\end{equation}
The $c_{ij}>0$ are the fixed ABHY constants for non-adjacent pairs $1\leq i<j\leq n-1$.  This is the ordered Koba-Nielsen integral written in positive coordinates~\cite{Koba:1969kh}.  It may equivalently be described using type-$\mathcal A$ binary variables $u_{ij}$ \cite{ArkaniHamedHeLamThomas2019,ArkaniHamedHeLamThomas2023,HeLiRamanZhang2020}.

Stringy canonical forms extend this example to integrals built from positive Laurent polynomials \cite{ArkaniHamedHeLamThomas2019}.  Under the usual convergence assumptions, if the $f_a$ are positive Laurent polynomials and the $c_a$ are positive weights, then
\begin{equation}
\begin{aligned}
	\lim_{\ap\to0^+} \ap^N \int_{\Rpos^N} \prod_{i=1}^N\dlog z_i\, \prod_a f_a(z)^{-\ap c_a} {=} &\int_{\R^N}\dd^N t\, e^{-\sum_a c_a\Trop(f_a)(t)}=\Omega\left(\bigoplus_a c_a\Newt(f_a)\right).
\end{aligned}
	\label{eq:intro-general-stringy-limit}
\end{equation}
This formula separates the two pieces of data used below: tropicalization turns the finite-$\ap$ integral into a Laplace integral, while the Newton polytopes of the positive Laurent polynomials determine the limiting canonical function. The polytope alone, however, does not determine the finite-$\ap$ integral, since different positive polynomials can share the same Newton polytope.  The question is whether a physical construction also selects the positive Laurent polynomials, not only their Newton polytope.

The positive-geometric study of cosmology began with cosmological polytopes and has since led to broader constructions, including cosmohedra and correlator polytopes \cite{Arkani-Hamed:2017fdk,Benincasa:2024leu,Arkani-Hamed:2024jbp, Ardila-Mantilla:2026cbo,Figueiredo:2025daa}. Recent work has also emphasized structural simplifications intrinsic to equal-time correlators, rather than inherited indirectly from wavefunction coefficients \cite{Chowdhury:2023arc,Arkani-Hamed:2025mce,Chowdhury:2026dwm}. Here we apply the mechanism above to fixed-graph in-in correlators, namely expectation values computed with the Schwinger--Keldysh, or in-in, prescription.  For the conformally coupled scalar theory considered below, the full-time representation (Section~\ref{sec:full-time-representation}) separates the fixed-graph correlator as $\corrG{=}\corrG_E\corrG_V$.  The factor $\corrG_V$ is a graph-dependent vertex-space Laplace integral, while $\corrG_E$ is the product of internal edge-energy propagators, equivalently written as an edge-space Laplace integral. The exponent of $\corrG_V$ contains the site terms $x_v|t_v|$ and the edge terms $y_e|t_u-t_v|$, which are the tropical limits of two positive Laurent binomials.  The corresponding Newton segments add up to the in-in zonotope $\Zono(G)$ introduced in \cite{Glew:2026von}.  The independent edge-space factor contributes a centered hypercube $H_\square(G)$.  Thus, in the fixed-graph cosmological setting, \eqref{eq:intro-general-stringy-limit} takes the following form, parallel to \eqref{eq:intro-phi3-abhy-string}:
\[
\begin{array}{cccc}
	&m_n^{\operatorname{Tr}(\phi^3)} =\Omega(\mathcal A_{n-3}) &\xleftarrow{\ \ap\to0^+\ }& \calI_n^{\operatorname{Tr}(\phi^3)}(\ap),\\[5pt]
	&\corrG =\Omega(\Zono(G))\times\Omega(H_\square(G)) &\xleftarrow{\ \ap\to0^+\ }& \calI_G^{\rm corr}(\ap).
\end{array}
\]
Keeping the same binomials before tropicalization gives the finite-$\ap$ stringy integral studied here.  Thus the physical time integral fixes both the polytope and its positive Laurent representation.  The edge-space factor is independent of the graph adjacency and deforms separately to a product of one-dimensional beta integrals.  We use ``stringy'' in this algebraic sense; no microscopic string-theory origin is assumed.

For comparison, the fixed-graph wavefunction leads to a different edge-space construction.  Let $\calB(G)$ be the collection of nonempty subsets $I\subseteq E(G)$ that induce connected subgraphs of $G$.  We define the $\mathcal G$-associahedron $\Asso(G)$ of $G$ as the graph associahedron $\operatorname{Asso}(L(G))$ of its line graph $L(G)$~\footnote{The line graph $L(G)$ has one vertex for each edge of $G$, and two vertices of $L(G)$ are adjacent whenever the corresponding edges of $G$ share an endpoint.}, with Minkowski realization \cite{CarrDevadoss2006GraphAssociahedra,Postnikov2009Permutohedra}
\begin{equation}
	\Asso(G):=\operatorname{Asso}(L(G)) = \bigoplus_{I\in\calB(G)}\lambda_I\,\Delta_I, \qquad \Delta_I:=\Conv\{\bfe_e:e\in I\}, \qquad \lambda_I>0.
	\label{eq:intro-g-associahedron}
\end{equation}
After factoring out the universal singleton poles $p_v$ and the total-energy pole $X_G$, the remaining tubing data are indexed by the nonempty connected proper edge sets $I\in\calB(G)$, $I\subsetneq E(G)$.  This is the tubing complex of $L(G)$, and $\Asso(G)$ has dimension $|E(G)|-1$.
\begin{figure}[htbp]
	\centering
	\includegraphics[width=0.9\linewidth]{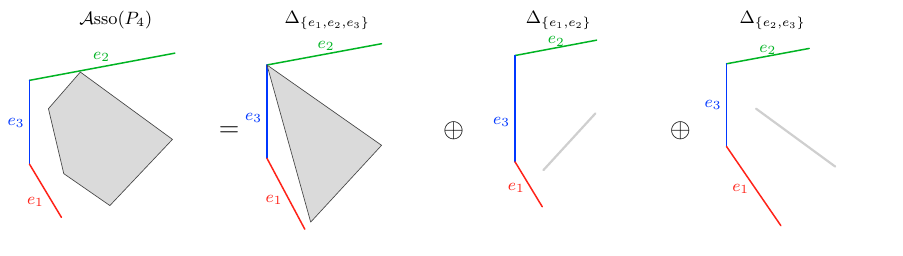}
	\caption{Minkowski-sum realization of the $\mathcal G$-associahedron $\Asso(P_4)$.}
	\label{fig:asso_minkowski}
\end{figure}
For example, when $G=P_4$, this Minkowski decomposition is illustrated in Figure~\ref{fig:asso_minkowski}.  To recover the physical wavefunction, the weights $\lambda_I$ of this ambient polytope must be pulled back to the linear combinations fixed by the site and edge energies.

There is a parallel edge-space realization of the fixed-graph correlator, referred to in this context as the graph correlahedron~\cite{Glew:2026von}.  For a tree graph $G$, this polytope is the graph cubeahedron of $L(G)$~\cite{DevadossHeathVipismakul2011}, which we call the $\mathcal G$-cubeahedron.  Appendix~\ref{app:tree-cubeahedron} gives its Minkowski-sum realization, the corresponding stringy integral, and its continuation to physical energies.  No such continuation is needed for the vertex-space in-in zonotope studied in the main text.  The $\mathcal G$-associahedral stringy integral and a fuller study of the $\mathcal G$-cubeahedral one are left to separate work.

For the graph-dependent vertex-space zonotope integral, we derive a uniform field-theory rational formula, together with exact parallel-edge reduction and edge-deletion identities at finite $\ap$.  At every partial energy pole, the residue factorizes into an internal block and shifted zonotope integrals for the complementary components; the internal block reduces to beta functions when the induced subgraph is a tree. For completeness, Appendix~\ref{app:critical-points} records the logarithmic critical equations and the generic reference degree of the associated affine divisor arrangement.  Whether this generic degree is retained under the constrained physical exponent specialization, and whether a corresponding critical-point pushforward exists, are left open.

The rest of the paper is organized as follows. Section~\ref{sec:review-correlators} reviews the fixed-graph correlator and fixes the energy variables.  Section~\ref{sec:stringy-zonotope} reads the full-time representation as a tropical integral, identifies the in-in zonotope, and constructs its stringy deformation.  It then derives the field-theory rational formula and gives basic examples.  Section~\ref{sec:finite-alpha} studies the finite-$\ap$ structure, and Section~\ref{sec:outlook} concludes with open problems.  Appendix~\ref{app:tree-cubeahedron} describes the tree-level cubeahedral alternative, while Appendix~\ref{app:critical-points} records the critical equations and generic reference degrees of the associated affine arrangement.

\section{Review of Fixed-Graph In-In Correlators}
\label{sec:review-correlators}

We first recall the fixed-graph contribution to equal-time in-in correlators and fix the graph and energy variables used throughout the paper.  In the Schwinger--Keldysh, or in-in, prescription \cite{Schwinger:1960qe,Keldysh:1964ud,Weinberg:2005vy,Arkani-Hamed:2017fdk}, expanding the two wavefunctionals perturbatively assigns every interaction vertex to one of the two time-contour branches; the Born rule then expresses the correlator in terms of wavefunction contributions.  We denote the connected contribution with fixed skeleton graph $G$ by $\corrG$.  We then pass to the full-time representation, which reorganizes the same fixed-graph contribution as the product of internal edge-energy propagators and a graph-dependent vertex-space Laplace integral.  These two factors will be deformed independently in Section~\ref{sec:stringy-zonotope}.

\subsection{Graph and energy variables}

Throughout the paper $G=(V,E)$ is a fixed finite connected loopless multigraph; parallel edges are distinct elements of $E(G)$.\footnote{Here $\textit{loopless}$ means that self-loops are excluded; graph cycles are allowed.}  We write $e=uv$ for an unoriented edge, $S\subseteq V(G)$ for a vertex subset, and $I\subseteq E(G)$ for an edge subset.  The corresponding vertex-induced and edge-induced subgraphs are $G[S]$ and $G[I]$; the latter has edge set $I$ and all endpoints of edges in $I$.  We also write $\Inc_G(v)$ for the set of edges incident to $v$, and $\pi_0(H)$ for the connected components of a graph $H$.  Unless stated otherwise, all integrals are first considered for $x_v>0$ and $y_e>0$.

\begin{figure}[htbp]
	\centering
	\includegraphics[width=0.4\linewidth]{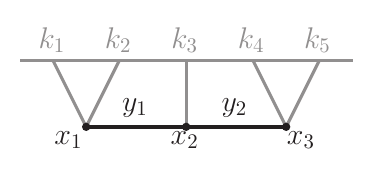}
	\caption{The $P_3$ graph as a skeleton graph of a five-point cubic Feynman graph.}
	\label{fig:skeletondiag}
\end{figure}

We regard $G$ as the skeleton graph of a Feynman diagram.  Each site $v\in V(G)$ carries a site energy $x_v$, the sum of the magnitudes of the external momenta entering that site.  For example, in Figure~\ref{fig:skeletondiag} one has $x_1=|\mathbf{k}_1|+|\mathbf{k}_2|$.  Each edge $e\in E(G)$ carries an edge energy $y_e$, the magnitude of the internal momentum flowing through that edge.  Momentum conservation in the same example gives $y_1=|\mathbf{k}_1+\mathbf{k}_2| =|\mathbf{k}_3+\mathbf{k}_4+\mathbf{k}_5|$.

For a subgraph $K\subseteq G$, let
\[
\partial_G(K):=\{uv\in E(G): |\{u,v\}\cap V(K)|=1\}, \qquad \partial_G(\ast):=\partial_G(G[\ast]).
\]
We use $p_v$ for the singleton tube variable and define the partial energy of a connected subgraph $K\subseteq G$ by
\begin{equation}
	p_v:=x_v+\sum_{e\in\Inc_G(v)}y_e, \qquad X_K:=\sum_{v\in V(K)}p_v-2\sum_{e\in E(K)}y_e.
	\label{eq:connected-cut-variable}
\end{equation}
The singularity at $X_K=0$ is a partial energy pole.  For a vertex-induced subgraph this reduces to
\begin{equation}
	X_S:=X_{G[S]} =\sum_{v\in S}x_v+\sum_{e\in\partial_G(S)}y_e\, .
	\label{eq:vertex-cut-variable}
\end{equation}
In particular, $X_{\{v\}}=p_v$ and $X_G=\sum_{v\in V(G)}x_v$ is the total energy.  If $K=G[I]$ is edge-induced, we also write $X_I:=X_{G[I]}$.  As sets of partial energy variables, the vertex-induced variables $X_S$ for subgraphs containing cycles are contained in the edge-induced variables $X_I$, by taking $I=E(G[S])$.  For correlators, the melonic simplification discussed in Section~\ref{sec:melonic} leaves only poles of the vertex-induced form $X_{G[S]}$; accordingly, the main correlator formulas use $X_S$ for the relevant partial energy poles.

We first recall the wavefunction of conformally coupled scalars, defined by the path integral~\cite{Arkani-Hamed:2025mce}
\begin{equation}
	\Psi[\Phi]= \int_{\varphi(t=-\infty)=0}^{\varphi(t=0)=\Phi}\mathcal D\varphi\, \exp\!\left[-i\int_{-\infty}^{0}d t\int_{-\infty}^{\infty} d^n x\, \mathcal S[\varphi]\right],
	\label{eq:wavefunction-path-integral}
\end{equation}
and encodes the quantum state on the $t=0$ slice.  Both the wavefunction and the correlator admit expansions in terms of Feynman graphs.  For the wavefunction, the coefficient $\Psi_G$ of a fixed skeleton graph $G$ is obtained from a single-branch time integral; after the standard Wick rotation, the conformally coupled scalar integrals become rational functions of the partial energies~\cite{Arkani-Hamed:2017fdk,Arkani-Hamed:2024jbp,Figueiredo:2025daa}. A convenient way to write this rational function is the tubing expansion~\cite{Glew:2025ugf}.

\begin{figure}[htbp]
	\centering
	\includegraphics[width=0.6\linewidth]{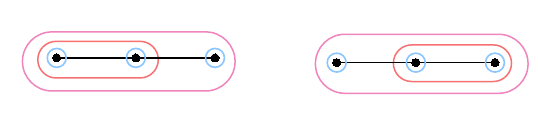}
	\caption{The two maximal tubings of the $P_3$ graph.  In each panel, the colored regions indicate the five compatible tubes.}
	\label{fig:maxtubing}
\end{figure}

A tube is a connected subgraph $K\subseteq G$. Two tubes $K_1,K_2$ are {\it compatible} if one contains the other, or if they share neither vertices nor edges.  A tubing is a pairwise compatible collection of tubes; a maximal tubing in the wavefunction expansion contains $|V(G)|+|E(G)|$ tube variables~\footnote{This terminology differs from the standard graph-associahedron convention: the nesting data we call tubes/tubings is closer to brackets/bracketings.  The edge-induced part of our convention becomes the usual tube/tubing language after passing to the line graph $L(G)$; see~\cite{CarrDevadoss2006GraphAssociahedra} for more details.}.  Let $\mathcal T_G$ be the set of maximal tubings, with each $K\in T$ carrying the partial energy $X_K$ of \eqref{eq:connected-cut-variable}.  Then
\begin{equation}
	\Psi_G=\sum_{T\in\mathcal T_G}\prod_{K\in T}\frac{1}{X_K}.
	\label{eq:wavefunction-tubing-expansion}
\end{equation}
After removing the common singleton and total-energy factors, the remaining tubing combinatorics define the $\calG$-associahedron.  For the path graph $P_3$ with vertices $v_1,v_2,v_3$,
\begin{equation}
	\Psi_{P_3}= \frac{1}{X_{\{v_1,v_2,v_3\}}p_{v_1}p_{v_2}p_{v_3}} \left(\frac{1}{X_{\{v_1,v_2\}}}+\frac{1}{X_{\{v_2,v_3\}}}\right),
	\label{eq:p3-wavefunction-example}
\end{equation}
where the total-energy pole and the single-vertex energy poles appear in every maximal tubing, since they are compatible with all other tubes. The two terms in parentheses correspond to the two maximal tubings shown in Figure~\ref{fig:maxtubing}.

The equal-time correlator is obtained from the wavefunction by the functional Born rule: one inserts a complete set of late-time field eigenstates, so that $|\Psi|^2$ defines the probability density on the space of boundary field configurations~\cite{Arkani-Hamed:2024jbp,Figueiredo:2025daa}.  These late-time correlators are the physical observables, with wavefunction coefficients entering as intermediate data through the Born rule.  Expanding this fixed-time average perturbatively gives a graph-level map from wavefunction coefficients to correlators~\cite{Glew:2025arc,Palma:2026qgn}; related all-order relations between cosmological correlators and wavefunction coefficients have been studied in~\cite{Stefanyszyn:2024wca}. For a fixed skeleton graph $G$, it can be organized as a sum over subsets $I\subseteq E(G)$ of internal edges.  Each edge in $I$ contributes a factor $(2y_e)^{-1}$, while deleting these edges leaves the wavefunction components $K\in\pi_0(G\setminus I)$. Thus
\begin{equation}
	\corrG=\sum_{I\subseteq E(G)} \frac{\prod_{K\in \pi_0(G\setminus I)}2\Psi_{K}} {\prod_{e\in I}2y_e}\,.
	\label{eq:born-rule-correlator}
\end{equation}
Here each component $K\in\pi_0(G\setminus I)$ is evaluated as a wavefunction graph with the tube variables, or equivalently the partial energies, inherited from $G$.

\subsection{The full-time representation}
\label{sec:full-time-representation}

For the conformally coupled scalar theory considered here, the equal-time in-in correlator can be evaluated, graph by graph, using a full-time Feynman-propagator representation~\cite{Donath:2024utn,Arkani-Hamed:2025mce,Glew:2026von}.  Equivalently, one may view this as the Wick-rotated time-ordered expression; since the external operators are inserted at equal time, the time-ordering of these external insertions is immaterial.  In graph variables, this representation assigns a factor $e^{-x_v|t_v|}$ to each site and the kernel \cite{Donath:2024utn,Glew:2026von}
\begin{equation}
	G_F(t_u,t_v;y_{uv})=\frac{1}{2y_{uv}}e^{-y_{uv}|t_u-t_v|}
	\label{eq:full-time-edge-kernel}
\end{equation}
to each internal edge $e=uv$.  Multiplying these factors and integrating one time variable for every site gives
\begin{equation}
	\corrG= \left(\prod_{e\in E(G)}\frac{1}{2y_e}\right) \int_{\R^{|V|}}e^{-\calS^V_G(t)}\,\dd t^V, \qquad \calS^V_G(t):= \sum_{v\in V(G)}x_v|t_v| +\sum_{uv\in E(G)}y_{uv}|t_u{-}t_v|.
	\label{eq:full-time-integral}
\end{equation}
Here $\dd t^V:=\prod_{v\in V(G)}\dd t_v$.  We will separate the graph-dependent vertex-space Laplace integral
\begin{equation}
	\corrG_V:= \int_{\R^{|V|}}e^{-\calS^V_G(t)}\,\dd t^V,
	\label{eq:full-time-integral-stripped}
\end{equation}
from the elementary edge-space factor.  The latter can itself be written as a product Laplace integral,
\begin{equation}
	\corrG_E:= \prod_{e\in E(G)}\frac{1}{2y_e} = \int_{\R^{|E|}}e^{-\calS^E_G(s)}\,\dd s^E, \qquad \calS^E_G(s):=4\sum_{e\in E(G)}y_e|s_e|,
	\label{eq:edge-space-laplace-factor}
\end{equation}
where $\dd s^E:=\prod_{e\in E(G)}\dd s_e$. Thus
\begin{equation}
	\corrG=\corrG_E\,\corrG_V.
	\label{eq:full-time-factorization}
\end{equation}

The Born-rule expression \eqref{eq:born-rule-correlator} and the full-time integral \eqref{eq:full-time-factorization} are two complementary organizations of the pole structure of the same fixed-graph correlator.  The Born rule makes the wavefunction tubing expansion manifest, while the full-time representation separates the elementary edge propagators and packages the graph-dependent part into $\corrG_V$.  The Born-rule rational form is naturally associated with the graph correlahedron; for a tree graph, Appendix~\ref{app:tree-cubeahedron} realizes this polytope as the $\mathcal G$-cubeahedron.  The full-time integral instead leads directly to the zonotope.  The rational form of $\corrG_V$ is discussed in Section~\ref{sec:augmented-chambers}.

\section{Tropical and Stringy In-In Correlator}
\label{sec:stringy-zonotope}

The full-time representation is already a tropical Laplace integral for the fixed-graph correlator.  Retaining the positive Laurent binomials whose tropical limits reproduce its absolute-value action gives its finite-$\ap$ stringy deformation.  We first construct the vertex-space and edge-space factors and identify their Newton polytopes.  We then use the zonotope geometry to obtain a uniform rational formula for $\corrG_V$ and work out basic examples.

\subsection{From the time integral to the zonotope and stringy integral}
We begin with the vertex-space action $\calS_G^V$ in \eqref{eq:full-time-integral}.  Following the general mechanism in \eqref{eq:intro-general-stringy-limit}, a stringy integral is naturally attached to a tropical Laplace integral once the latter is written as the tropicalization of positive Laurent factors.  The point here is not to first guess a polytope and then search for a finite-$\ap$ deformation; the relevant Laurent factors are already dictated by the tropical form of the time integral. For a positive Laurent polynomial
\[
F({\mathbf z})=\sum_{\mathbf m}c_{\mathbf m}{\mathbf z}^{\mathbf m}, \qquad c_{\mathbf m}>0,
\]
we use the max-plus tropicalization
\begin{equation}
	\Trop(F)({\mathbf t}):=\max_{\mathbf m}(\mathbf m\cdot \mathbf t).
	\label{eq:tropicalization-definition}
\end{equation}
The positive coefficients do not affect this tropical function. For example,
\begin{equation}
	\Trop(1+2z_1+3z_1z_2+z_1z_3^{-1})(\mathbf t) =\max(0,t_1,t_1+t_2,t_1-t_3).
	\label{eq:tropicalization-example}
\end{equation}

Equivalently, after the logarithmic substitution $z_i{=}e^{t_i/\ap}$, one has $\ap\log F(e^{\mathbf t/\ap})\to\Trop(F)(\mathbf t)$ as $\ap{\to}0^+$.  The two absolute-value terms in $\calS_G^V$ arise in this way from Laurent binomials:
\begin{equation}
	\begin{aligned}
		\ap\log(z_v+z_v^{-1})&\longrightarrow \max(-t_v,t_v)=|t_v|,\\
		\ap\log\!\left(\frac{z_u}{z_v}+\frac{z_v}{z_u}\right) &\longrightarrow \max(t_v{-}t_u,t_u{-}t_v)=|t_u-t_v|\,.
	\end{aligned}
	\label{eq:direct-tropical-dictionary}
\end{equation}

The Newton polytopes of the two Laurent binomials in \eqref{eq:direct-tropical-dictionary} are, respectively, $[-\bfe_v,\bfe_v]$ and $[-(\bfe_u-\bfe_v),\bfe_u-\bfe_v]$.  Their weighted Minkowski sum is the in-in zonotope introduced in \cite{Glew:2026von}:
\begin{equation}
	\Zono(G):= \bigoplus_{v\in V(G)}x_v[-\bfe_v,\bfe_v] \,\oplus\, \bigoplus_{uv\in E(G)} y_{uv}[-(\bfe_u{-}\bfe_v),\,\bfe_u{-}\bfe_v].
	\label{eq:inin-zonotope-short}
\end{equation}
This gives the vertex-space stringy integral directly from the time integral:
\begin{equation}
	\calI^{\zono}_{G}(\ap)= \ap^{|V|} \int_{\Rpos^{|V|}}\prod_{v\in V(G)}\dlog z_v\, \prod_{v\in V(G)}(z_v+z_v^{-1})^{-\ap x_v} \prod_{uv\in E(G)} \left(\frac{z_u}{z_v}+\frac{z_v}{z_u}\right)^{-\ap y_{uv}}.
	\label{eq:zonoint}
\end{equation}
After the substitution $z_v{=}e^{t_v/\ap}$, dominated convergence in the positive-energy region gives
\begin{equation}
	\lim_{\ap\to0^+}\calI_G^{\zono}(\ap)=\corrG_V.
	\label{eq:zono-field-theory-limit}
\end{equation}
The positive Laurent representation in \eqref{eq:zonoint} is therefore inherited from the full-time action, rather than chosen from the Newton polytope alone.

\begin{figure}[htbp]
	\centering
	\includegraphics[width=0.8\linewidth]{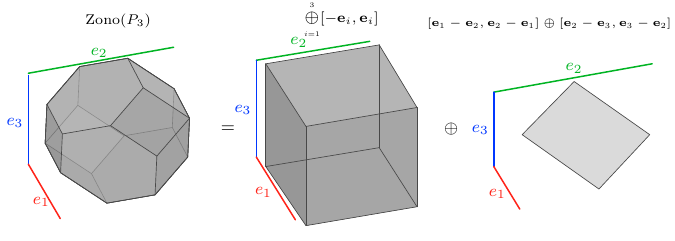}
	\caption{Minkowski-sum realization of $\Zono(P_3)$ from its site and edge segments.}
	\label{fig:zono_minkowski}
\end{figure}

The same tropical action is the support function of $\Zono(G)$.  Its domains of linearity form the normal fan of $\Zono(G)$, equivalently the normal fan of the Newton polytope of the Laurent product in \eqref{eq:zonoint}.  The unit sublevel set is the polar body $\Zono(G)^\vee$, and radial integration gives~\cite{ArkaniHamedHeLamThomas2019}
\begin{equation}
	\corrG_V=\Omega_G^{\zono}:=\Omega(\Zono(G)) =\int_{\R^{|V|}}e^{-\calS_G^V(t)}\,\dd t^V =|V(G)|!\,\Vol(\Zono(G)^\vee).
	\label{eq:zono-normalized-dual-volume}
\end{equation}
Figure~\ref{fig:zono_minkowski} illustrates this Minkowski-sum realization for the in-in zonotope of path graph $P_3$.

The edge-space factor $\corrG_E$ in \eqref{eq:edge-space-laplace-factor} is independent of the graph adjacency. Its stringy deformation is the product of one-dimensional beta integrals
\begin{equation}
	\calI^{\square}_{G}(\ap)= \ap^{|E(G)|}\int_{\Rpos^{|E(G)|}}\prod_{e\in E(G)}\dlog z_e\, \prod_{e\in E(G)}(z_e+z_e^{-1})^{-4\ap y_e} =\prod_{e\in E(G)}\frac{\ap}{2}B(2\ap y_e,2\ap y_e),
	\label{eq:hypercube-stringy-factor}
\end{equation}
with $\lim_{\ap\to0^+}\calI_G^{\square}(\ap) =\prod_e(2y_e)^{-1}$.  The Newton segments of the edge factors form the hypercube
\begin{equation}
	H_{\square}(G):= \bigoplus_{e\in E(G)}4y_e[-\bfe_e,\bfe_e].
\end{equation}
The complete stringy correlator is
\begin{equation}
	\calI_G^{\rm corr}(\ap) :=\calI_G^{\square}(\ap)\,\calI_G^{\zono}(\ap), \qquad \lim_{\ap\to0^+}\calI_G^{\rm corr}(\ap)=\corrG.
	\label{eq:fullzonoint}
\end{equation}
The Newton polytope controlling the field-theory limit is therefore $H_\square(G)\times\Zono(G)$ in dimension $|E|+|V|$.  The hypercube records the elementary propagator normalization, whereas all dependence on the adjacency of $G$ sits in the $\Zono(G)$ factor.  Since $\calI_G^{\square}$ is elementary, the rest of the paper focuses on the vertex-space pair $\corrG_V$ and $\calI_G^{\zono}(\ap)$.

For later use, the change of variables $z_v=e^{\tau_v}$, with $\tau_v\in\R$, writes the latter as a real integral better suited to the analysis of finite-$\ap$ behavior:
\begin{equation}
	\calI_G^{\zono}(\ap) =\ap^{|V(G)|}\int_{\R^{V(G)}}\prod_{v\in V(G)}\dd\tau_v\, \prod_{v\in V(G)}(2\cosh\tau_v)^{-\ap x_v} \prod_{e=uv\in E(G)}(2\cosh(\tau_u{-}\tau_v))^{-\ap y_e}.
	\label{eq:stringy-time-integral}
\end{equation}

\subsection{Rational form from augmented-graph chambers}
\label{sec:augmented-chambers}

The support-function picture turns the full-time integral directly into a general rational form~\eqref{eq:general-zono-chamber-formula} for $\corrG_V$.  Indeed, $\calS_G^V(t)$ is the support function of $\Zono(G)$, whose normal fan records the maximal domains on which the action is linear.  After a simplicial refinement, the integral over each such domain is elementary.  We now make this chamber decomposition explicit for any connected loopless graph, including graphs with cycles.

To treat the site and edge terms uniformly, define the augmented graph
\begin{equation}
	\widehat G:=G\vee\{0\},
\end{equation}
as the complete union of $G$ with a single auxiliary vertex $0$.  By the complete union $G_1\vee G_2$, we mean the disjoint union with all original edges retained and an additional edge between every vertex of $G_1$ and every vertex of $G_2$.  Thus $0$ is adjacent to every vertex of $G$.  Assign the edges of $\widehat G$ the energies
\begin{equation}
	w_{0v}=x_v, \qquad w_e=y_e \quad(e\in E(G)).
\end{equation}

\begin{figure}[htbp]
	\centering
	\includegraphics[width=0.3\linewidth]{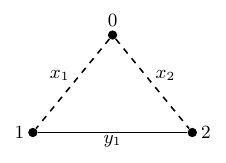}
	\caption{ The augmented graph $\widehat P_2$.  The solid edge $12$ belongs to the original graph $P_2$ and carries the energy $y_{1}$, while the dashed edges $01$ and $02$ carry $x_1$ and $x_2$, respectively.
	}
	\label{fig:augmented-graph}
\end{figure}

The vertex-space action can then be written uniformly as
\begin{equation}
	\widehat\calS_G^V(t;t_0) = \sum_{uv\in E(\widehat G)} w_{uv}|t_u-t_v|, \qquad \calS_G^V(t)=\widehat\calS_G^V(t;0).
	\label{eq:augmented-graph-action}
\end{equation}

The weighted graphical zonotope of $\widehat G$ is identified with $\Zono(G)$ under $[\bfe_v{-}\bfe_0]\mapsto \bfe_v$; dually, fixing $t_0{=}0$ reduces its support function to $\calS_G^V(t)$. For every nonempty proper subset $A\subsetneq V(\widehat G)$, write $\bar A:=V(\widehat G)\setminus A$ and define its cut energy by
\begin{equation}
	\widehat{X}_A := \sum_{e\in\partial_{\widehat G}(A)}w_e.
	\label{eq:augmented-cut-energy}
\end{equation}
In terms of the partial energies of $G$, this is equivalently
\begin{equation}
	\widehat{X}_A=
	\begin{cases}
		\displaystyle\sum_{K\in \pi_0(G[A])}X_K, &0\notin A,\\[7pt]
		\displaystyle\sum_{K\in\pi_0(G[\bar A])}X_K, &0\in A.
	\end{cases}
	\label{eq:augmented-cut-component-formula}
\end{equation}
Since a cut is unchanged upon exchanging its two sides, we have
\begin{equation}
	\widehat{X}_A=\widehat{X}_{\bar A},\qquad \bar A=V(\widehat G)\setminus A.
\end{equation}

Let $n=|V(G)|$.  The graphical arrangement of $\widehat G$, formed by the hyperplanes $t_u=t_v$ for $uv\in E(\widehat G)$, has the axis through $\mathbf 0$ and $(1,\ldots,1)$ as a common boundary of all its cones. Projection along this axis onto $t_0=0$ by
\[
(t_0,t_1,\ldots,t_n)\longmapsto(t_1-t_0,\ldots,t_n-t_0)
\]
identifies its quotient fan with the normal fan of $\Zono(G)$ in $\mathbb R^{V(G)}$.

For the integration, consider the graph-independent arrangement of all hyperplanes $t_i=t_j$, with $0\leq i<j\leq n$.  These hyperplanes divide $\mathbb R^{n+1}$ into $(n+1)!$ ordering cones.  For each $\sigma=(\sigma_0,\ldots,\sigma_n)\in S_{n+1}$, define
\begin{equation}
	\operatorname{Cone}(\sigma) := \left\{ \mathbf t\in\mathbb R^{n+1}: t_{\sigma_0}\leq t_{\sigma_1}\leq\cdots\leq t_{\sigma_n} \right\}.
\end{equation}
These cones have disjoint interiors.  Their intersections $\operatorname{Cone}(\sigma)\cap \{\mathbf t\in\mathbb R^{n+1}:t_0=0\}$ are pointed $n$-dimensional cones with common apex at the origin and cover $\mathbb R^{V(G)}$.  The cones are paired by central inversion: if $\sigma^{\mathrm{rev}}=(\sigma_n,\ldots,\sigma_0)$, then
\begin{equation}
	\operatorname{Cone}(\sigma^{\mathrm{rev}}) =-\operatorname{Cone}(\sigma).
\end{equation}
This universal decomposition can be finer than the maximal linearity domains of $\widehat\calS_G^V$.  Indeed, if $ij\notin E(\widehat G)$, crossing the hyperplane $t_i{=}t_j$ does not change the action, so the two adjacent ordering cones belong to the same maximal domain.  This refinement does not affect the integral, since their intersections with the slice still partition the integration domain up to measure-zero boundaries.

Introduce the prefix sets and adjacent differences
\begin{equation}
	A_k^\sigma:=\{\sigma_0,\ldots,\sigma_{k-1}\}, \qquad \delta_k:=t_{\sigma_k}-t_{\sigma_{k-1}}\geq0, \qquad k=1,\ldots,n.
\end{equation}
The $\delta_k$ form coordinates on $\operatorname{Cone}(\sigma)\cap \{\mathbf t\in\mathbb R^{n+1}:t_0=0\}$ with unit Jacobian determinant.

\begin{figure}[htbp]
	\centering
	\includegraphics[width=0.8\linewidth]{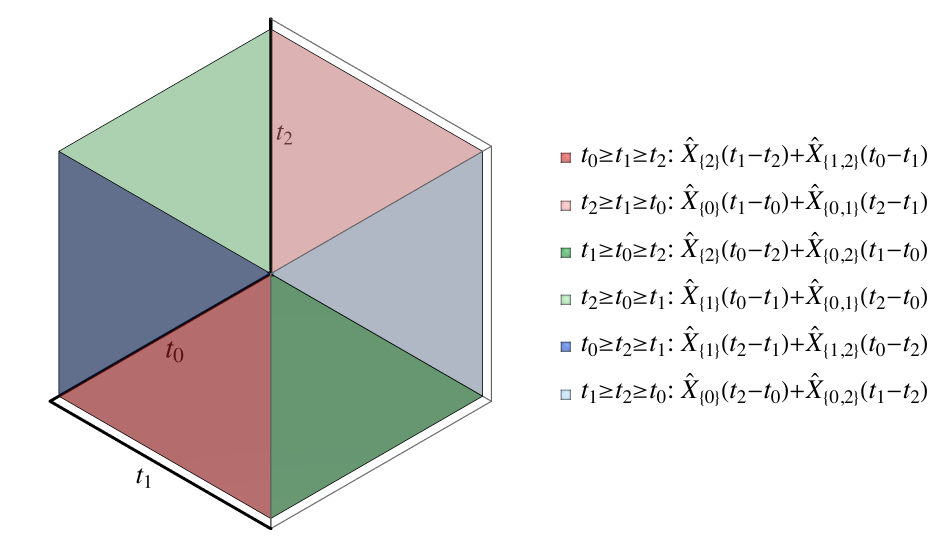}
	\caption{The six ordering cones for the augmented graph $\widehat P_2$, viewed along their common axis in $(t_0,t_1,t_2)$-space.  Each colored cone corresponds to an ordering $(t_{\sigma_0},t_{\sigma_1},t_{\sigma_2})$, with $\sigma\in S_3$, and the legend gives the corresponding local linear form of the action $\widehat\calS_{P_2}^V$ in terms of the cut energies $\widehat X_A$.  Projection along the same axis onto $t_0=0$ gives the normal fan of $\Zono(P_2)$.}
	\label{fig:augmented-decomposition}
\end{figure}

An edge $uv$ contributes to $\delta_k$ in $|t_u-t_v|$ precisely when its endpoints lie on opposite sides of the prefix cut $A_k^\sigma$.  Hence, on the cone associated with $\sigma$, the action is
\begin{equation}
	\left.\widehat\calS_G^V(t;t_0)\right|_{\operatorname{Cone}(\sigma)\cap \{\mathbf t\in\mathbb R^{n+1}:t_0=0\}} = \sum_{k=1}^n \widehat{X}_{A_k^\sigma}\delta_k.
	\label{eq:local-augmented-action}
\end{equation}
The simplest example, $G=P_1$, is
\begin{equation}
	\widehat\calS_{P_1}^V(t;t_0)=x_1|t_1-t_0|=
	\begin{cases}
		x_1(t_1-t_0),& \operatorname{Cone}((0,1))\\
		x_1(t_0-t_1),& \operatorname{Cone}((1,0))
	\end{cases}\,,
\end{equation}
The corresponding three-coordinate picture for $\widehat P_2$ is shown in Figure~\ref{fig:augmented-decomposition}.

The contribution of $\operatorname{Cone}(\sigma)\cap\{t_0=0\}$ is therefore the factorized Laplace integral
\begin{equation}
	\int_{\operatorname{Cone}(\sigma)\cap \{\mathbf t\in\mathbb R^{n+1}:t_0=0\}}\dd^n t\, e^{-\calS_G^V(t)} = \int_{\mathbb R_{\geq0}^n}\dd^n \delta\, \exp\!\left(-\sum_{k=1}^n\widehat X_{A_k^\sigma}\delta_k\right) = \prod_{k=1}^n\frac{1}{\widehat X_{A_k^\sigma}}.
\end{equation}

For the centrally symmetric cone associated with $\sigma^{\mathrm{rev}}$, the prefix cuts obey
\begin{equation}
	A_k^{\sigma^{\mathrm{rev}}} = V(\widehat G)\setminus A_{n+1-k}^{\sigma}.
\end{equation}
The identity $\widehat X_A=\widehat X_{\bar A}$ therefore matches the denominators in reverse order.  Thus every pair of centrally symmetric cones makes equal contributions to $\Omega_G^{\zono}$.

Figure~\ref{fig:augmented-decomposition} shows the six ordering cones for $\widehat P_2$, viewed along this common axis. Because $\widehat P_2$ is complete on $\{0,1,2\}$, its graphical arrangement contains all three hyperplanes $t_i=t_j$, so the six ordering cones are exactly the six maximal linearity domains.  Their projections along this axis onto $t_0=0$ form the normal fan of $\Zono(P_2)$.  The zonotope itself is shown later in Figure~\ref{fig:zono_minkowski_P2}.

Summing the cone contributions gives the uniform formula
\begin{equation}
        \corrG_V=\Omega_G^{\zono} = \sum_{\sigma\in S_{n+1}} \prod_{k=1}^n \frac{1}{\widehat{X}_{A_k^\sigma}}.
	\label{eq:general-zono-chamber-formula}
\end{equation}

The chamber formula uses all prefix cuts of the braid fan, but the actual facets of the zonotope have a simpler graph-theoretic description.
\begin{proposition}
	\label{prop:zono-connected-facets}
	The pairs of opposite facets of $\Zono(G)$ are indexed by the nonempty connected vertex subsets $S\subseteq V(G)$.  The corresponding facet variable is the partial energy $X_S$.
\end{proposition}
\begin{proof}
Since $\Zono(G)$ is a centered realization of the weighted graphical zonotope of $\widehat G$, its pairs of opposite facets are indexed by cuts $A\mid\bar A$ for which both $\widehat G[A]$ and $\widehat G[\bar A]$ are connected ~\cite[Sec.~1.3]{PadrolPilaudPoullot2025}.  Choosing the side not containing $0$ and denoting it by $S\subseteq V(G)$, the other side is automatically connected because it contains the universal vertex $0$.  Hence the opposite facet pairs of $\Zono(G)$ are indexed by nonempty subsets $S\subseteq V(G)$ for which $G[S]$ is connected, with corresponding facet variable $X_S$.
\end{proof}
Consequently, a denominator associated with a disconnected prefix set may appear in an individual term of~\eqref{eq:general-zono-chamber-formula}, since the braid fan can refine the graphical fan, but it is not a facet pole and cancels from the full chamber sum.

The same example makes the cut identifications and central pairing concrete. For $\widehat P_2$, the six nontrivial subsets form three complementary pairs, with cut energies
\begin{equation}
	\begin{aligned}
		\widehat X_{\{0\}}=\widehat X_{\{1,2\}} &=X_{\{1,2\}}=x_1+x_2,\\
		\widehat X_{\{1\}}=\widehat X_{\{0,2\}} &=X_{\{1\}}=x_1+y_{1},\\
		\widehat X_{\{2\}}=\widehat X_{\{0,1\}} &=X_{\{2\}}=x_2+y_{1}.
	\end{aligned}
\end{equation}
Thus the six augmented cut energies reduce to the three partial energies $X_{12}$, $X_1$, and $X_2$ of the original graph.  Since every nonempty subset of $P_2$ is connected, these label the three pairs of opposite facets of $\Zono(P_2)$.  Listing the six permutations of $V(\widehat P_2)=\{0,1,2\}$ in the order used in Figure~\ref{fig:augmented-decomposition} gives
\begin{align}
	\Omega_{P_2}^{\zono} &= \frac{1}{\widehat X_{\{2\}}\widehat X_{\{1,2\}}} + \frac{1}{\widehat X_{\{0\}}\widehat X_{\{0,1\}}} + \frac{1}{\widehat X_{\{2\}}\widehat X_{\{0,2\}}}+ \frac{1}{\widehat X_{\{1\}}\widehat X_{\{0,1\}}} + \frac{1}{\widehat X_{\{1\}}\widehat X_{\{1,2\}}} + \frac{1}{\widehat X_{\{0\}}\widehat X_{\{0,2\}}} \notag\\
	&= 2\left( \frac{1}{X_2X_{12}} + \frac{1}{X_2X_1} + \frac{1}{X_1X_{12}} \right)\,.
	\label{eq:P2-augmented-chamber-example}
\end{align}
The second equality groups the six ordering-cone contributions into three centrally symmetric pairs, whose members are equal by $\widehat X_A=\widehat X_{\bar A}$.

The chamber representation~\eqref{eq:general-zono-chamber-formula} of $\Omega_G^{\zono}$ is obtained from the ordering-cone fan, which is a simplicial refinement of the normal fan of $\Zono(G)$.  Individual chamber terms may therefore contain $\widehat X_A$ associated with disconnected subsets $A$.  These denominators correspond to internal rays introduced by the refinement rather than to facets of $\Zono(G)$.  When the ordering cones contained in each cone of the normal fan are combined, the auxiliary denominators cancel, leaving only the facet variables $X_S$ identified in Proposition~\ref{prop:zono-connected-facets}.  The resulting expression is algebraically equivalent to the rational form organized directly in terms of the $X_S$ poles in~\cite{Glew:2025arc}.

\subsection{Examples}
\label{sec:examples}
\subsubsection{The one-vertex and one-edge graphs}

For the single-vertex graph $P_1$, the stringy in-in correlator is the stringy integral for the centered interval,
\[
\calI_{P_1}^{\zono}(\ap) = \ap\int_0^\infty (z_1+z_1^{-1})^{-\ap x_1}\,\dlog z_1 = \frac{\ap}{2} B\left(\frac{\ap x_1}{2},\frac{\ap x_1}{2}\right) \xrightarrow[]{\ap\to0^+}\frac{2}{x_1}.
\]
The basic site segment $[-\bfe,\bfe]$ therefore has normalized dual length $\Omega_{P_1}^{\zono}=2/x_1$; the factor of two comes from the centered normalization.

\begin{figure}[htbp]
	\centering
	\includegraphics[width=0.9\linewidth]{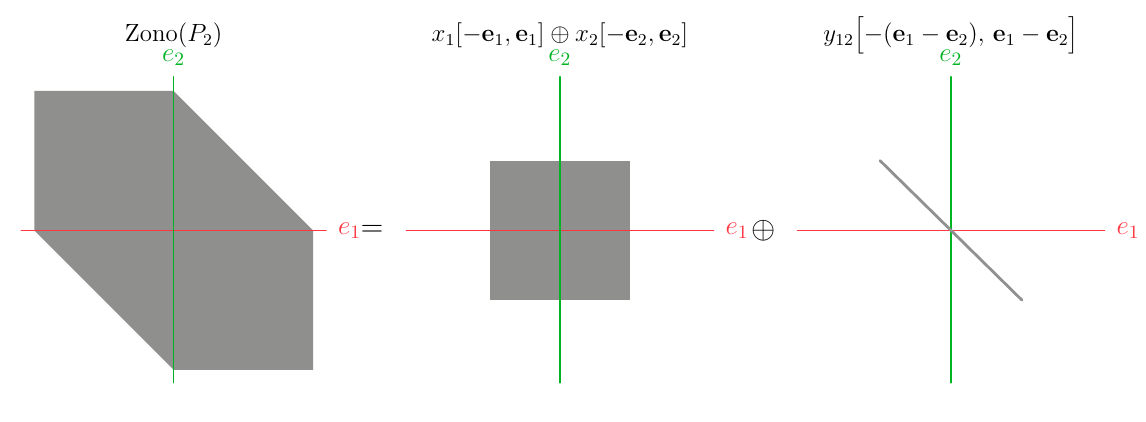}
	\caption{Minkowski decomposition of $\Zono(P_2)$.  The two site segments form the rectangle in the center, whose Minkowski sum with the edge segment on the right gives the zonotope on the left.}
	\label{fig:zono_minkowski_P2}
\end{figure}

For the one-edge path graph $P_2$, the Laurent form is
\[
\calI_{P_2}^{\zono}(\ap) = \ap^2 \int_{\Rpos^2}\dlog z_1\,\dlog z_2\, (z_1+z_1^{-1})^{-\ap x_1} (z_2+z_2^{-1})^{-\ap x_2} \left(\frac{z_1}{z_2}+\frac{z_2}{z_1}\right)^{-\ap y_1}.
\]
The corresponding in-in zonotope is a three-segment zonotope as shown in Figure~\ref{fig:zono_minkowski_P2}.

Applying the augmented-graph chamber formula \eqref{eq:general-zono-chamber-formula}, whose $P_2$ evaluation was given explicitly in \eqref{eq:P2-augmented-chamber-example}, and fully expanding the partial energies in terms of $x_1,x_2$, and $y_1$, gives the normalized dual volume, or equivalently the stripped full-time correlator,
\begin{equation}
	\Omega_{P_2}^{\zono}(x_1,x_2,y_{1}) = 4\,\frac{x_1+x_2+y_{1}} {(x_1+x_2)(x_1+y_{1})(x_2+y_{1})}.
	\label{eq:one-edge-zono-rational}
\end{equation}
Restoring the elementary edge propagator gives the complete correlator
\begin{equation}
	\langle P_2\rangle=\frac{1}{2y_{1}}\,\Omega_{P_2}^{\zono}\,.
\end{equation}
Setting $y_{1}=0$ gives
\[
\Omega_{P_2}^{\zono}(x_1,x_2,y_{1})\xrightarrow{y_{1}=0} \,\Omega_{P_1}^{\zono}(x_1)\Omega_{P_1}^{\zono}(x_2)=\frac{4}{x_1x_2},
\]
which is the disconnected split \eqref{eq:disconnected-edge-deletion}.

\subsubsection{Path graphs \texorpdfstring{$P_n$}{P(n)}}

For the path graph $P_n$, $n\geq1$, on vertices $1,\ldots,n$, with edge energy $y_i$ on $i\,i{+}1$, the augmented graph $\widehat P_n$ defined above is the fan graph obtained by adjoining the vertex $0$ to every site of the path.  Equation~\eqref{eq:general-zono-chamber-formula} therefore gives an explicit sum over $(n+1)!$ ordering cones.  For $n=1$, the chamber formula gives $2/x_1$; for $n=2$, it gives \eqref{eq:one-edge-zono-rational}.

For $P_3$, the augmented-graph chamber formula specializes to
\begin{equation}
	\Omega_{P_3}^{\zono} = \sum_{\sigma\in S_{4}} \frac{1}{\widehat{X}_{\{\sigma_0\}}\widehat{X}_{\{\sigma_0,\sigma_1\}}\widehat{X}_{\{\sigma_0,\sigma_1,\sigma_2\}}}.
	\label{eq:general-zono-chamber-formula-P3}
\end{equation}
The disconnected subset $\{1,3\}$ introduces the auxiliary denominator $\widehat X_{\{1,3\}}=X_{\{1\}}+X_{\{3\}}$, which does not correspond to a physical partial-energy pole.  Upon grouping the chamber contributions, the terms involving this denominator contain the factor
\begin{equation}
	\frac{1}{X_{\{1\}}+X_{\{3\}}} \left( \frac{1}{X_{\{1\}}}+\frac{1}{X_{\{3\}}} \right) = \frac{1}{X_{\{1\}}X_{\{3\}}}.
\end{equation}
Thus the factor $X_{\{1\}}+X_{\{3\}}$ cancels.  Rewriting every $\widehat X_A$ in terms of the partial energies and simplifying the auxiliary denominators gives
\begin{equation}
	\begin{aligned}
		\Omega_{P_3}^{\zono} =2\Bigg( &\frac{1}{X_{\{1\}}X_{\{1,2\}}X_{\{1,2,3\}}} +\frac{1}{X_{\{1\}}X_{\{1,2\}}X_{\{3\}}} +\frac{1}{X_{\{1\}}X_{\{1,2,3\}}X_{\{3\}}} +\frac{1}{X_{\{1\}}X_{\{2\}}X_{\{2,3\}}} \\
		&+\frac{1}{X_{\{1\}}X_{\{2\}}X_{\{3\}}} +\frac{1}{X_{\{1\}}X_{\{2,3\}}X_{\{3\}}} +\frac{1}{X_{\{1,2\}}X_{\{1,2,3\}}X_{\{2\}}} +\frac{1}{X_{\{1,2\}}X_{\{2\}}X_{\{3\}}} \\
		&+\frac{1}{X_{\{1,2,3\}}X_{\{2\}}X_{\{2,3\}}} +\frac{1}{X_{\{1,2,3\}}X_{\{2,3\}}X_{\{3\}}} \Bigg).
	\end{aligned}
	\label{eq:P3-path-rational-form}
\end{equation}
The auxiliary denominator $\widehat X_{\{1,3\}}$ has therefore disappeared: \eqref{eq:P3-path-rational-form} is organized entirely in terms of physical partial-energy poles.  This also illustrates the non-uniqueness of rational organizations discussed above.

If deleting an edge separates the path into two paths, $P_L$ and $P_R$, then
\[
\Omega_{P_n}^{\zono}\big|_{y_e=0} = \Omega_{P_L}^{\zono}\Omega_{P_R}^{\zono}.
\]

\subsubsection{Star graphs
	\texorpdfstring{$K_{1,n}$}{K(1,n)} and cycle graphs \texorpdfstring{$C_n$}{C(n)}}

For the star $K_{1,n}$, write $c$ for the central vertex and $1,\ldots,n$ for the leaves, with edge energy $y_i$ on $ci$.  For $A\subseteq\{1,\ldots,n\}$, the cut energies of the augmented graph are
\begin{equation}
	\begin{aligned}
		\widehat X_A &=\sum_{i\in A}(x_i+y_i), &&A\ne\varnothing, \\
		\widehat X_{\{c\}\cup A} &=x_c+\sum_{i\in A}x_i+\sum_{j\notin A}y_j, \\
		\widehat X_{\{0\}\cup A} &=x_c+\sum_{i\notin A}x_i+\sum_{j\in A}y_j, \\
		\widehat X_{\{0,c\}\cup A} &=\sum_{i\notin A}(x_i+y_i), &&A\ne\{1,\ldots,n\}.
	\end{aligned}
\end{equation}
The last two expressions are the complements of the first two.  A leaf-only subset with $|A|\geq2$ is disconnected, so the denominator $\widehat X_A=\sum_{i\in A}X_i$ is auxiliary rather than a physical pole and cancels after the chamber contributions are summed.  The first instance is $K_{1,2}\simeq P_3$, where the two-leaf subset produces precisely the analogue of $\widehat X_{\{1,3\}}=X_1+X_3$ above.

Every star edge is a bridge, namely, an edge whose deletion disconnects the graph.  Consequently, setting $y_i=0$ separates the leaf $i$ and gives
\begin{equation}
	\left.\Omega_{K_{1,n}}^{\zono}\right|_{y_i=0} = \Omega_{P_1}^{\zono}(x_i)\, \Omega_{K_{1,n-1}}^{\zono}.
\end{equation}

By contrast, no edge of a cycle $C_n$ is a bridge.  For $n\geq4$, a vertex subset may consist of several disjoint cyclic arcs, say $S=A_1\sqcup\cdots\sqcup A_r$.  Such a subset produces $\widehat X_S=\sum_{a=1}^r X_{A_a}$, again an auxiliary chamber denominator that cancels from the final rational form.  The triangle $C_3$ is exceptional. Its augmented graph is $\widehat C_3=K_4$, and the chamber formula specializes to
\begin{equation}
	\Omega_{C_3}^{\zono} = \sum_{\sigma\in S_4} \frac{1}{ \widehat X_{\{\sigma_0\}} \widehat X_{\{\sigma_0,\sigma_1\}} \widehat X_{\{\sigma_0,\sigma_1,\sigma_2\}}}.
	\label{eq:C3-chamber-formula}
\end{equation}
Every nonempty subset of $C_3$ is connected.  Thus, for each cut of $\widehat C_3$, the side not containing $0$ labels a physical partial energy $X_S$; unlike for $P_3$, no auxiliary denominator associated with a disconnected subset appears.  Pairing centrally symmetric chambers and rewriting the result in terms of these partial energies gives
\begin{equation}
	\begin{aligned}
		\Omega_{C_3}^{\zono} = &2\Bigg( \frac{1}{X_{\{1\}}X_{\{1,2\}}X_{\{1,2,3\}}} +\frac{1}{X_{\{1\}}X_{\{1,2\}}X_{\{3\}}} +\frac{1}{X_{\{1\}}X_{\{1,2,3\}}X_{\{1,3\}}} \\
		&+\frac{1}{X_{\{1\}}X_{\{1,3\}}X_{\{2\}}} +\frac{1}{X_{\{1\}}X_{\{2\}}X_{\{2,3\}}} +\frac{1}{X_{\{1\}}X_{\{2,3\}}X_{\{3\}}} \\
		&+\frac{1}{X_{\{1,2\}}X_{\{1,2,3\}}X_{\{2\}}} +\frac{1}{X_{\{1,2\}}X_{\{2\}}X_{\{3\}}} +\frac{1}{X_{\{1,2,3\}}X_{\{1,3\}}X_{\{3\}}} \\
		&+\frac{1}{X_{\{1,2,3\}}X_{\{2\}}X_{\{2,3\}}} +\frac{1}{X_{\{1,2,3\}}X_{\{2,3\}}X_{\{3\}}} +\frac{1}{X_{\{1,3\}}X_{\{2\}}X_{\{3\}}} \Bigg).
	\end{aligned}
	\label{eq:C3-zono-rational-form}
\end{equation}
It is manifest from this expression that only the partial-energy poles $X_S=X_{C_3[S]}$ of vertex-induced subgraphs appear. Assigning $y_1,y_2,y_3$ to the edges $12,23,31$, respectively, and setting $y_3=0$ deletes the edge $31$ and produces a path rather than a product:
\begin{equation}
	\left.\Omega_{C_3}^{\zono}\right|_{y_3=0} = \Omega_{P_3}^{\zono}\,.
\end{equation}
Thus a cycle already distinguishes edge deletion from disconnected factorization.  The finite-$\ap$ deletion identities for these graph families are discussed in Section~\ref{sec:finite-alpha}; their generic arrangement reference degrees are recorded separately in Appendix~\ref{app:critical-points}.

\section{Finite-\texorpdfstring{$\alpha'$}{alpha'} Structure}
\label{sec:finite-alpha}

We now study the finite-$\ap$ behavior of the zonotopal stringy integral. Having separated off the edge-space beta functions in Section~\ref{sec:stringy-zonotope}, we focus here on $\calI_G^{\zono}(\ap)$, which encodes the nontrivial fixed-graph dependence of the correlator. Reducing each parallel-edge bundle to a single edge carrying its total energy, we may work with a weighted simple skeleton. There are then two types of factorization channels: with the pole already stripped off together with the edge-space prefactor, $\calI_G^{\zono}(\ap)$ reduces to the edge-deleted graph, splitting into a product over its connected components; at a partial-energy pole $X_S=0$, the residue of $\calI_G^{\zono}(\ap)$ factorizes into an internal block for $G[S]$ and shifted zonotopal integrals for the components of its complement. When $G[S]$ is a tree, the internal block reduces to a product of beta functions, while for $G[S]$ with cycles it remains a coupled integral. The partial-energy channel can have descendant poles at the even levels $X_S=-2m/\ap$, while the odd levels are absent.

\subsection{Parallel-edge reduction and melonic simplification}\label{sec:melonic}

\begin{figure}[htbp]
	\centering
	\includegraphics[width=0.75\linewidth]{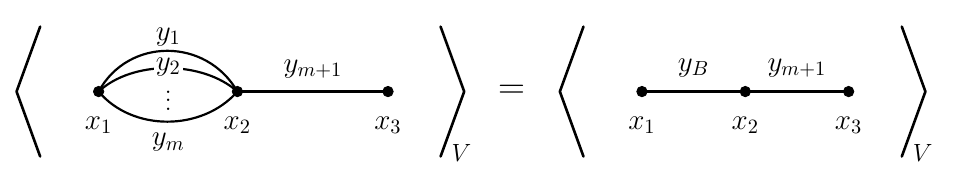}
	\caption{Melonic simplification for a three-site multigraph.}
	\label{fig:Melonic}
\end{figure}

At the field-theory level, correlators of multigraphs obey a melonic simplification~\cite{Arkani-Hamed:2025mce,Donath:2024utn}.  Suppose two vertices $a,b$ of a multigraph $G_{\rm multi}$ are connected by a parallel-edge bundle
\[
B=\{e_1,\ldots,e_m\}.
\]
Let $G_{\rm simple}$ be the simple skeleton obtained by replacing this bundle with a single edge $(a,b)$ of energy $y_B:=\sum_{e\in B}y_e$.  Then
\begin{equation}
	\langle G_{\rm multi}\rangle_V=\langle G_{\rm simple}\rangle_V \implies \langle G_{\rm multi}\rangle= \frac{2y_B}{\prod_{e\in B}2y_e}\, \langle G_{\rm simple}\rangle\,.
	\label{eq:raw-melonic-born-coefficient}
\end{equation}
The replacement of the parallel-edge bundle by a single edge of total energy $y_B$ is illustrated in Figure~\ref{fig:Melonic}.

After the propagator poles are stripped off, melonic reduction leaves only the connected vertex-induced poles of $G$.  For graphs with cycles this is a much smaller set than the corresponding wavefunction pole set.

The zonotopal stringy integral obeys the same reduction at finite $\ap$. All edges in $B$ contribute the same hyperbolic factor, so their energies appear only through their sum:
\begin{equation}
	\prod_{e\in B} (2\cosh(\tau_a-\tau_b))^{-\ap y_e} = (2\cosh(\tau_a-\tau_b))^{-\ap y_B}.
	\label{eq:parallel-edge-reduction}
\end{equation}
It follows that the zonotope integral for a multigraph is the same as the zonotope integral for its simple skeleton, with each parallel bundle replaced by a single edge of total weight $y_B$:
\begin{equation}
	\calI_{G_{\rm multi}}^{\zono}(\ap) = \calI_{G_{\rm simple}}^{\zono}(\ap).
	\label{eq:melonic-reduction}
\end{equation}
Here and below, the edge $(a,b)$ of $G_{\rm simple}$ carries the total energy $y_B$.  The identity holds at finite $\ap$, not only in the field-theory limit.  For the complete stringy correlator, the only remaining effect comes from the elementary edge-space hypercube factor:
\begin{equation}
	\calI_{G_{\rm multi}}^{\rm corr}(\ap) = \calI_{G_{\rm multi}}^{\square}(\ap)\, \calI_{G_{\rm simple}}^{\zono}(\ap) = \frac{ \displaystyle\prod_{e\in B}\frac{\ap}{2}B(2\ap y_e,2\ap y_e) }{ \displaystyle\frac{\ap}{2}B(2\ap y_B,2\ap y_B) }\, \calI_{G_{\rm simple}}^{\rm corr}(\ap).
\end{equation}
Thus the exact melonic simplification is a vertex-space phenomenon: the zonotopal factor only remembers whether two vertices are adjacent and the total weight of the corresponding direction, while the number of parallel edges is recorded separately by the elementary edge-space factor.

\subsection{Factorizations on energy poles}
\subsubsection{Edge energy poles}

At simultaneous edge-energy poles, a fixed-graph correlator factorizes into the correlators of the connected components obtained by deleting the corresponding edges~\cite{Figueiredo:2025daa}.  More precisely, let $I\subseteq E(G)$, and write
\[
\pi_0(G\setminus I)=\{K_1,K_2,\ldots,K_r\},
\]
where each $K_i$ is allowed to be a single-vertex graph.  At the field-theory level, this gives
\begin{equation}
	\Res_{y_e=0\, (e\in I)}\corrG =\left(\Res_{y_e=0\, (e\in I)}\corrG_E\right)\left(\corrG_V|_{y_e=0\, (e\in I)}\right)= \frac{1}{2^{|I|}} \prod_{K_i\in\pi_0(G\setminus I)}\langle K_i\rangle.
	\label{eq:field-theory-edge-factorization}
\end{equation}
The first equality separates the universal edge poles from the regular vertex-space factor: the residue is entirely carried by $\corrG_E$, while $\corrG_V$ is evaluated at $y_e=0$ for $e\in I$.  The factor $2^{-|I|}$ comes from the explicit propagators $1/(2y_e)$. Equation~\eqref{eq:field-theory-edge-factorization} is immediate from either the Born-rule expression~\eqref{eq:born-rule-correlator} or the full-time integral~\eqref{eq:full-time-integral}.

The complete stringy correlator has the analogous finite-$\ap$ factorization
\[
\Res_{y_e=0\, (e\in I)}\calI_G^{\rm corr}(\ap) = \frac{1}{2^{|I|}} \prod_{K_i\in\pi_0(G\setminus I)}\calI_{K_i}^{\rm corr}(\ap).
\]
Indeed, the elementary edge-space factor contains all the edge poles, with
\[
\frac{\ap}{2}B(2\ap y_e,2\ap y_e) = \frac{1}{2y_e}+O(y_e) \qquad (y_e\to0),
\]
while $\calI_G^{\zono}(\ap)$ is regular at $y_e=0$.  Thus the only graph-dependent statement to establish is the behavior of the zonotopal factor at that locus.  In the hyperbolic representation, each edge $e=uv$ contributes $(2\cosh(\tau_u-\tau_v))^{-\ap y_e}$, which becomes $1$ when $y_e=0$. Therefore
\begin{equation}
	\calI_G^{\zono}(\ap)\big|_{y_e=0\, (e\in I)} = \calI_{G\setminus I}^{\zono}(\ap),
	\label{eq:edge-deletion}
\end{equation}
with the same site energies and all remaining edge energies.  If deleting the edges in $I$ disconnects the graph, the integration variables separate between its connected components, giving
\begin{equation}
	\calI_G^{\zono}(\ap)\big|_{y_e=0\, (e\in I)} = \prod_{K_i\in \pi_0(G\setminus I)}\calI_{K_i}^{\zono}(\ap).
	\label{eq:disconnected-edge-deletion}
\end{equation}
Combining this identity with the elementary edge-space residues gives the complete stringy factorization above, whose field-theory limit is \eqref{eq:field-theory-edge-factorization}.

\subsubsection{Partial energy poles}
\begin{figure}[htbp]
	\centering
	\includegraphics[width=0.6\linewidth]{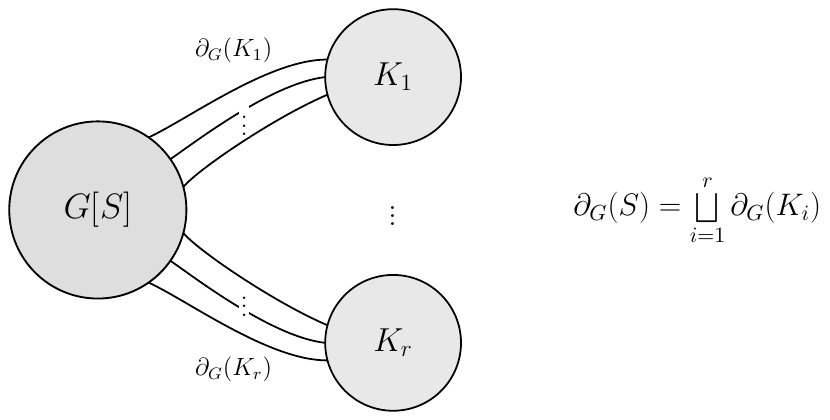}
	\caption{Decomposition of the cut boundary. Let $K_1,\ldots,K_r$ be the connected components of $G\setminus S$. The set of edges joining $G[S]$ to $K_i$ is the boundary $\partial_G(K_i)$, and $\partial_G(S)=\bigsqcup_{i=1}^{r}\partial_G(K_i)$.}
	\label{fig:connected-cut-configuration}
\end{figure}

The partial energy singularities of cosmological correlators have been studied from several complementary perspectives~\cite{Baumann:2021fxj,Chowdhury:2026upp}. Unlike the edge-energy pole at $y_e=0$, the partial-energy channel produces $X_S$ poles directly in $\calI_G^{\zono}(\ap)$. These poles arise from the boundary regions of the vertex-space integral in which the time variables associated with $S$ are sent collectively to $\tau=\pm\infty$, while their relative separations remain finite. To be precise, let $S\subseteq V(G)$ be nonempty, with $G[S]$ connected, and write
\[
\pi_0(G\setminus S)=\{K_1,\ldots,K_r\},
\]
for the connected components of its complement, as illustrated in Figure~\ref{fig:connected-cut-configuration}.

For a given cut $S$ with connected $G[S]$, we first focus on the pole at $X_S=0$. Since it comes from translating all vertices in $S$ together to $\tau=\pm\infty$, its residue naturally factorizes into an internal factor $\calC_S(\ap)$ for $G[S]$ and a shifted zonotope block $\calJ_{K_j}^{S}(\ap)$ for each component $K_j$ of its complement:
\begin{equation}
	\operatorname*{Res}_{X_S=0}\calI_G^{\zono}(\ap) = 2\,\mathcal C_S(\ap) \prod_{j=1}^r\mathcal J_{K_j}^{S}(\ap).
	\label{eq:zono-general-connected-cut-factorization-short}
\end{equation}

For each complementary component $K_j$, the contribution $\calJ_{K_j}^{S}(\ap)$ is given by the zonotope integral for $K_j$ with vertex-dependent shifts $\eta^S_v$ induced by the cut edges:
\begin{align}
	\mathcal J_{K_j}^{S}(\ap) &:=\ap^{|V(K_j)|} \int_{\R^{V(K_j)}} \prod_{v\in V(K_j)}\dd\tau_v\, \prod_{v\in V(K_j)} (2\cosh\tau_v)^{-\ap x_v} \nonumber\\
	&\quad\times \prod_{uv\in E(K_j)} (2\cosh(\tau_u-\tau_v))^{-\ap y_{uv}} \exp\!\left[\ap\sum_{v\in V(K_j)}\eta^{S}_v\tau_v\right].
	\label{eq:zono-shifted-component-short}
\end{align}
Here the shifts $\eta^S_v$ are given by the total energy of the cut edges incident to each vertex $v$:
\begin{equation}
	\eta^S_v:=\sum_{e\in\partial_G(S)\cap\Inc_G(v)}y_e.
\end{equation}

For the internal subgraph $G[S]$, we may choose an arbitrary reference vertex $a_\star\in S$ and set $\tau_{a_\star}=0$. The internal factor $\calC_S(\ap)$ is then given by
\begin{align}
	\mathcal C_S(\ap):=\ap^{|S|-1}\int_{\R^{S\setminus\{a_\star\}}}&\prod_{v\in S\setminus\{a_\star\}}\dd \tau_v\, \exp\!\left[-\ap\sum_{v\in S}(x_v+\eta_v^S)\tau_v\right]\nonumber\\
    \times&\prod_{uv\in E(G[S])}\!(2\cosh(\tau_u{-}\tau_v))^{-\ap y_{uv}}.
	\label{eq:zono-connected-cut-internal-block-short}
\end{align}
The integral is understood by meromorphic continuation to $X_S=0$. The result is independent of the choice of $a_\star$: changing the anchor shifts all anchored coordinates by a common constant, whose coefficient in the exponent is $X_S$ and therefore vanishes on the pole.

To see explicitly how this factorization arises from the boundary regions of the integral in which $\tau_v\to\pm\infty$ simultaneously for $v\in S$, we set $T=\tau_{a_\star}$ and introduce the variable substitution
\[
	\tau_v=T+\sigma_v, \qquad v\in S, \qquad \sigma_{a_\star}=0,
\]
while leaving those $\tau_p$, $p\notin S$ unchanged. This isolates the common translation of $\tau_v$, $v\in S$, with the two boundary regions corresponding to $T\to\pm\infty$. Let us first consider $T\to+\infty$. Collecting all $T$-dependent factors in the integrand, we find
\begin{equation}
\begin{aligned}
	&\prod_{v\in S} (2\cosh(T+\sigma_v))^{-\ap x_v} \prod_{\substack{e=vp\in\partial_G(S)\\v\in S,\ p\notin S}} (2\cosh(T+\sigma_v-\tau_p))^{-\ap y_e}\\
	&\qquad=e^{-\ap X_ST} \exp\!\left[-\ap\sum_{v\in S}(x_v+\eta_v^S)\sigma_v\right] \exp\!\left[\ap\sum_{p\notin S}\eta_p^S\tau_p\right] \left(1+\mathcal O(e^{-2T})\right).
\end{aligned}
\end{equation}

After inserting this asymptotic form into $\calI_G^{\zono}$, we isolate the collective $T$-integral and collect the remaining factors into $\calC_S$ and $\calJ_{K_j}^S$. The full integral near $X_S=0$ is then
\begin{align}
	\calI_G^{\zono}(\ap)={}&\ap\int_{T_0}^{\infty}\dd T\,e^{-\ap X_ST}\,\calC_S(\ap)\prod_{j=1}^{r}\calJ_{K_j}^{S}(\ap)\nonumber\\
    &+(\text{singularity from }T\to-\infty)+(\text{terms regular at }X_S=0).
\end{align}

Since the integral is invariant under the simultaneous reflection $\tau_v\mapsto-\tau_v$, the two boundary regions give identical singular contributions. Thus, near $X_S=0$,
\begin{align}
	\calI_G^{\zono}(\ap) &=2\ap\int_{T_0}^{\infty}\dd T\,e^{-\ap X_ST}\,\calC_S(\ap)\prod_{j=1}^{r}\calJ_{K_j}^{S}(\ap)+(\text{terms regular at }X_S=0)\nonumber\\
	&=\frac{2}{X_S}\calC_S(\ap)\prod_{j=1}^{r}\calJ_{K_j}^{S}(\ap)+(\text{terms regular at }X_S=0).
\end{align}
Taking the residue at $X_S=0$ gives the aforementioned factorization~\eqref{eq:zono-general-connected-cut-factorization-short}.

The internal block $\calC_S$ is much simpler than the zonotopal stringy integral for $G[S]$. This follows from the asymptotic $T$-translation invariance of the large-$|T|$ region responsible for the residue: the $T$-dependent $\cosh$ factors reduce to their leading exponentials, whose combined dependence on a collective time translation of $S$ vanishes on $X_S=0$.

When $G[S]$ is a tree, its $|S|-1$ edge differences are independent, so the integral in \eqref{eq:zono-connected-cut-internal-block-short} splits into a product of one-dimensional beta integrals. For each $e\in E(G[S])$, let $D_e\subset S$ and $S\setminus D_e$ be the two vertex sets obtained by deleting $e$. Since the beta function is symmetric, the choice of $D_e$ is immaterial. If the edge difference is oriented towards $D_e$, its linear coefficient is $q_e=\sum_{v\in D_e}(x_v+\eta_v^S)=X_{D_e}-y_e$. Using
\begin{equation}
	\ap\int_{\R}e^{-\ap q r}(2\cosh r)^{-\ap y}\,\dd r =\frac{\ap}{2} B\!\left(\frac{\ap}{2}(y+q),\frac{\ap}{2}(y-q)\right)
	\label{eq:one-dimensional-cosh-beta}
\end{equation}
and $X_{D_e}+X_{S\setminus D_e}=2y_e$ on $X_S=0$, we obtain
\begin{equation}
	\mathcal C_S(\ap)=\mathcal F_S(\ap) := \prod_{e\in E(G[S])} \frac{\ap}{2} B\!\left( \frac{\ap}{2}X_{D_e}, \frac{\ap}{2}X_{S\setminus D_e} \right)\bigg|_{X_S=0}.
	\label{eq:zono-tree-cut-beta-block-short}
\end{equation}
Substituting $\mathcal C_S=\mathcal F_S$ into \eqref{eq:zono-general-connected-cut-factorization-short} gives
\begin{equation}
	\operatorname*{Res}_{X_S=0}\calI_G^{\zono}(\ap) = 2 \left( \prod_{j=1}^r \mathcal J_{K_j}^{S}(\ap) \right) \mathcal F_S(\ap) \qquad \text{when $G[S]$ is a tree}.
	\label{eq:zono-general-tree-tube-factorization-short}
\end{equation}
The shifts record how the cut is attached to the complement. For a cut in a tree, each complementary component is attached through one edge. A proper arc in a cycle has two boundary attachments, while a general cut may have several.

If $G[S]$ contains cycles, its edge differences obey one constraint for each independent cycle. The internal factor $\calC_S$ is then a coupled $(|S|-1)$-dimensional integral rather than a product of independent beta functions, while the general residue formula~\eqref{eq:zono-general-connected-cut-factorization-short} remains unchanged.

In the field-theory limit, the internal block becomes the fixed-graph flat-space block associated with $G[S]$, which we denote by $\mathcal A_{G[S]}$:
\begin{equation}
\begin{aligned}
	\mathcal A_{G[S]} &:= \lim_{\ap\to0^+}\mathcal C_S(\ap) \\
	&= \int_{\R^{S\setminus\{a_\star\}}} \prod_{v\in S\setminus\{a_\star\}}\dd\tau_v\, \exp\!\Bigg[ -\sum_{v\in S}(x_v+\eta_v^S)\tau_v -\sum_{uv\in E(G[S])}y_{uv}|\tau_u-\tau_v| \Bigg] \bigg|_{X_S=0},
\end{aligned}
\label{eq:zono-internal-amplitude-ft}
\end{equation}
with $\tau_{a_\star}=0$.  For a subgraph containing cycles, $\mathcal A_{G[S]}$ is understood as the corresponding $\ell_0$-integrated amplitude~\cite{Arkani-Hamed:2025mce}.

In the same limit, each shifted complementary block becomes a shifted version of the stripped full-time correlator
\begin{equation}
\begin{aligned}
	\Omega_{K_j}^{S,\zono} &:= \lim_{\ap\to0^+}\mathcal J_{K_j}^{S}(\ap) \\
	&= \int_{\R^{V(K_j)}} \prod_{v\in V(K_j)}\dd\tau_v\, \exp\!\Bigg[ -\sum_{v\in V(K_j)}x_v|\tau_v| -\sum_{uv\in E(K_j)}y_{uv}|\tau_u{-}\tau_v| +\sum_{v\in V(K_j)}\eta_v^S\tau_v \Bigg].
\end{aligned}
\label{eq:zono-shifted-omega-block-short}
\end{equation}
When the tilt vanishes, this reduces to the ordinary $\Omega_{K_j}^{\zono}$.

Taking the field-theory limit of \eqref{eq:zono-general-connected-cut-factorization-short} therefore gives, for any connected $G[S]$,
\begin{equation}
	\operatorname*{Res}_{X_S=0}\Omega_G^{\zono} = 2\,\mathcal A_{G[S]} \prod_{j=1}^r\Omega_{K_j}^{S,\zono}.
	\label{eq:zono-general-connected-cut-factorization-ft}
\end{equation}
For a tree subgraph, the internal amplitude evaluates explicitly to
\begin{equation}
	\mathcal A_{G[S]} = \prod_{e\in E(G[S])} \frac{2y_e}{X_{D_e}X_{S\setminus D_e}} \bigg|_{X_S=0},
	\label{eq:zono-tree-internal-amplitude-ft}
\end{equation}
and
\begin{equation}
	\operatorname*{Res}_{X_S=0}\Omega_G^{\zono} = 2 \left( \prod_{j=1}^r \Omega_{K_j}^{S,\zono} \right) \left( \prod_{e\in E(G[S])} \frac{2y_e}{X_{D_e}X_{S\setminus D_e}} \right), \qquad G[S]\ \text{a tree}.
	\label{eq:zono-general-tree-tube-factorization-ft-short}
\end{equation}

\begin{figure}[t]
	\centering
	\includegraphics[width=\linewidth]{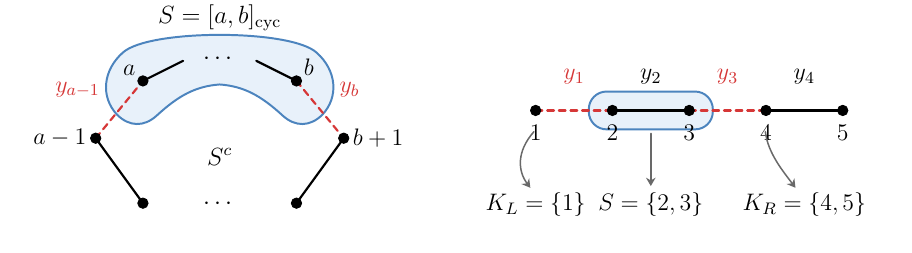}
	\caption{Connected-cut configurations at finite $\alpha'$. Left: a proper arc $S=[a,b]_{\mathrm{cyc}}$ of $C_n$. Right: the $X_{\{2,3\}}$ cut of $P_5$.}
	\label{fig:connected-cut-example}
\end{figure}

\paragraph{Proper arcs of a cycle.}
Let $G=C_n$, label the edge $(i,i{+}1)$ by $y_i$ cyclically, and take a proper cyclic arc
\[
S=[a,b]_{\rm cyc}\subsetneq V(C_n).
\]
The channel variable and the two boundary edges are
\begin{equation}
	X_S=\sum_{i\in S}x_i+y_{a-1}+y_b, \qquad (a{-}1,a),\ (b,b{+}1).
	\label{eq:cycle-proper-arc-channel}
\end{equation}
This proper-arc configuration is shown in the left panel of Figure~\ref{fig:connected-cut-example}.

Both induced graphs $C_n[S]$ and $C_n[S^c]$ are paths.  The internal block is therefore the beta product $\mathcal F_S$ in \eqref{eq:zono-tree-cut-beta-block-short}, while the complement carries a two-sided tilt,
\begin{equation}
	\begin{aligned}
		\mathcal J_{S^c}^{LR}(\ap) &:= \ap^{|S^c|} \int_{\R^{S^c}} \prod_{i\in S^c}\dd\tau_i\, \prod_{i\in S^c}(2\cosh\tau_i)^{-\ap x_i} \prod_{e=ij\in E(C_n[S^c])} (2\cosh(\tau_i-\tau_j))^{-\ap y_e} \\&\times\exp\!\left[\ap\bigl(y_b\tau_{b+1} +y_{a-1}\tau_{a-1}\bigr)\right].
		\label{eq:cycle-two-sided-tilted-block}
	\end{aligned}
\end{equation}
Specializing \eqref{eq:zono-general-tree-tube-factorization-short} gives
\begin{equation}
	\operatorname*{Res}_{X_S=0}\calI_{C_n}^{\zono}(\ap) =2\,\mathcal F_S(\ap)\,\mathcal J_{S^c}^{LR}(\ap).
	\label{eq:cycle-proper-arc-factorization}
\end{equation}
The internal block remains a beta product, but the complementary path is tilted at both endpoints.  This is the first difference from a cut of a tree.

By contrast, in the total-energy channel $S=V(C_n)$, the internal graph is the whole cycle and there is no complementary block.  The general formula gives
\begin{equation}
	\operatorname*{Res}_{X_{V(C_n)}=0}\calI_{C_n}^{\zono}(\ap) = 2\,\mathcal C_{V(C_n)}(\ap).
	\label{eq:cycle-total-energy-connected-cut}
\end{equation}
The edge differences in $\mathcal C_{V(C_n)}$ obey one cycle constraint, so this is a coupled cyclic block rather than a beta product.

\paragraph{Example: the $X_{\{2,3\}}$ cut of $P_5$.}
Consider the five-site path $P_5$, with $y_i:=y_{i,i+1}$, and choose $S=\{2,3\}$, as shown in the right panel of Figure~\ref{fig:connected-cut-example}.  The relevant channel variables are
\[
X_S=x_2+x_3+y_1+y_3, \qquad p_2=x_2+y_1+y_2, \qquad p_3=x_3+y_2+y_3,
\]
so that $p_2+p_3=2y_2$ on $X_S=0$.  The complement has two connected components,
\[
\pi_0(P_5\setminus S)=\{K_L,K_R\}, \qquad K_L=P_5[\{1\}], \qquad K_R=P_5[\{4,5\}],
\]
and the two cut edges induce the shifts $y_1$ and $y_3$.  The left component is the tilted one-vertex block
\[
\mathcal J_{K_L}^{S}(\ap) = \frac{\ap}{2} B\!\left( \frac{\ap}{2}(x_1+y_1), \frac{\ap}{2}(x_1-y_1) \right),
\]
whereas the right component is the shifted two-site block
\[
\begin{aligned}
	\mathcal J_{K_R}^{S}(\ap) ={} \ap^2 \int_{\R^2}\dd\tau_4\,\dd\tau_5\, (2\cosh\tau_4)^{-\ap x_4} (2\cosh\tau_5)^{-\ap x_5} (2\cosh(\tau_4-\tau_5))^{-\ap y_4} \exp(\ap y_3\tau_4).
\end{aligned}
\]
Choosing $D_{(2,3)}=\{2\}$, the residue becomes
\begin{equation}
	\operatorname*{Res}_{X_{\{2,3\}}=0} \calI_{P_5}^{\zono}(\ap) = 2\, \mathcal J_{K_L}^{S}(\ap)\, \mathcal J_{K_R}^{S}(\ap)\, \frac{\ap}{2} B\!\left( \frac{\ap p_2}{2}, \frac{\ap p_3}{2} \right)\bigg|_{X_{\{2,3\}}=0}.
	\label{eq:p5-two-site-connected-cut}
\end{equation}
Thus the internal edge $y_2$ supplies the beta-function block, while the cut edges $y_1$ and $y_3$ appear as exponential tilts of the complementary components.

Taking the field-theory limit, the two complementary blocks become the shifted $\Omega^{\zono}$ blocks
\begin{align}
	\Omega_{K_L}^{S,\zono} &:= \lim_{\ap\to0^+}\mathcal J_{K_L}^{S}(\ap) = \int_{\R}\dd \tau_1\, e^{-x_1|\tau_1|+y_1\tau_1}= \frac{2x_1}{x_1^2-y_1^2},
	\label{eq:p5-left-tilted-block-ft}
	\\
	\Omega_{K_R}^{S,\zono} &:= \lim_{\ap\to0^+}\mathcal J_{K_R}^{S}(\ap)= \int_{\R^2}\dd \tau_4\,\dd \tau_5\, \exp\!\left[ -x_4|\tau_4| -x_5|\tau_5| -y_4|\tau_4-\tau_5| +y_3\tau_4 \right].
	\label{eq:p5-right-tilted-block-ft}
\end{align}
The latter shifted two-site block has the explicit rational form
\begin{equation}
	\begin{aligned}
		\Omega_{K_R}^{S,\zono} = \sum_{\epsilon=\pm1} \Bigg[ \frac{1}{x_4+x_5+\epsilon y_3} \left( \frac{1}{x_4+y_4+\epsilon y_3} + \frac{1}{x_5+y_4} \right) + \frac{1} {(x_4+y_4+\epsilon y_3)(x_5+y_4)} \Bigg].
	\end{aligned}
	\label{eq:p5-right-tilted-block-rational}
\end{equation}
The beta-function block associated with the internal edge has the limit
\begin{equation}
	\begin{aligned}
		\lim_{\ap\to0^+} \frac{\ap}{2} B\left( \frac{\ap p_2}{2}, \frac{\ap p_3}{2} \right)= \frac{p_2+p_3}{p_2p_3} = \frac{2y_2}{p_2p_3} \qquad \text{on }X_{\{2,3\}}=0,
	\end{aligned}
	\label{eq:p5-beta-block-ft}
\end{equation}
where we used $p_2+p_3=X_{\{2,3\}}+2y_2$. Consequently, the field-theory residue factorizes as
\begin{equation}
	\operatorname*{Res}_{X_{\{2,3\}}=0} \Omega_{P_5}^{\zono} = 2\, \Omega_{K_L}^{S,\zono}\, \Omega_{K_R}^{S,\zono}\, \frac{2y_2}{p_2p_3} \bigg|_{X_{\{2,3\}}=0}.
	\label{eq:p5-two-site-connected-cut-ft}
\end{equation}
This reproduces the sign-shift structure of the tree-level correlator factorization of~\cite{Arkani-Hamed:2025mce}.

\subsection{Even descendant towers}
As for general stringy canonical forms, the meromorphic structure is closely tied to the facet structure of the underlying Newton polytope \cite{ArkaniHamedHeLamThomas2019,HeLiRamanZhang2020}. The edge-space factor $\calI_G^{\square}(\ap)$ is a product of one-variable beta functions, as shown in \eqref{eq:hypercube-stringy-factor}, so its meromorphic structure is determined edge by edge.  We therefore focus on the graph-dependent zonotopal factor~\eqref{eq:stringy-time-integral}. The possible pole channels correspond to pairs of opposite facets of $\Zono(G)$, which are labeled by nonempty connected subsets $S\subseteq V(G)$.  For each such $S$, the meromorphic continuation of $\calI_G^{\zono}(\ap)$ can have poles only in the evenly spaced tower
\begin{equation}
	X_S=-\frac{2m}{\ap}, \qquad m=0,1,2,\ldots .
	\label{eq:even-tower-pole}
\end{equation}

The $m{=}0$ member is the partial-energy pole $X_S{=}0$, while $m{>}0$ gives its finite-$\ap$ descendants.

To derive this result for a fixed $S$, translate all its vertices together while keeping their relative positions and the variables outside $S$ fixed. With $X_S$ defined in \eqref{eq:vertex-cut-variable}, choose a reference vertex $v_\star\in S$ and define
\[
T:=\tau_{v_\star},\qquad \sigma_v:=\tau_v-T\quad(v\in S),
\]
so that $\sigma_{v_\star}=0$ and $\tau_v=T+\sigma_v$ for $v\in S$; the variables $\tau_w$, with $w\notin S$, are left unchanged.  When $S=V(G)$, there are no outside variables or cut-edge factors.

For edges internal to $G[S]$, the differences $\tau_u{-}\tau_v=\sigma_u{-}\sigma_v$ are independent of $T$; the same is true of all factors supported entirely on the complement of $S$.  Thus the $T$-dependence comes only from the site factors in $S$ and the edges crossing $\partial_G(S)$.  To expand these factors near the two endpoints, introduce
\[
\lambda_+=e^{-T}\quad(T\to+\infty),\qquad \lambda_-=e^{T}\quad(T\to-\infty).
\]
For either a site factor or a cut-edge factor, its $T$-dependent part takes the common form
\[
(2\cosh(T+d))^{-\ap c} = \lambda_\pm^{\ap c}e^{\mp\ap c d} \bigl(1+\lambda_\pm^2e^{\mp2d}\bigr)^{-\ap c},
\]
where $(c,d)=(x_v,\sigma_v)$ for a site $v\in S$, and $(c,d)=(y_e,\sigma_v-\tau_w)$ for a cut edge $e=vw$, with $v\in S$ and $w\notin S$.  The last factor is a power series in $\lambda_\pm^2$.  Multiplying the leading powers over all such factors gives
\[
\lambda_\pm^{\ap(\sum_{v\in S}x_v+ \sum_{e\in\partial_G(S)}y_e)} =\lambda_\pm^{\ap X_S}.
\]
Consequently, the full product of hyperbolic factors has the endpoint expansion
\begin{equation}
	\prod_v(2\cosh\tau_v)^{-\ap x_v} \prod_{e=uv\in E}(2\cosh(\tau_u-\tau_v))^{-\ap y_e} = \lambda_\pm^{\ap X_S} \sum_{m\geq0}\lambda_\pm^{2m}R_{2m}^{\pm}.
	\label{eq:cut-tower}
\end{equation}
Here $R_{2m}^{\pm}$ are independent of $T$ and collect the remaining factors together with the coefficients of the binomial expansions; they depend on the remaining integration variables, $\ap$, and the energy parameters.  Since $\dd T=\mp\dd\lambda_\pm/\lambda_\pm$, termwise integration near either endpoint reduces to
\[
\int_0^\epsilon\dd\lambda_\pm\, \lambda_\pm^{\ap X_S+2m-1} = \frac{\epsilon^{\ap X_S+2m}}{\ap X_S+2m}.
\]
The meromorphic continuation of $\calI_G^{\zono}(\ap)$ can therefore have poles only at the locations in \eqref{eq:even-tower-pole}.  In particular, the absence of odd powers in \eqref{eq:cut-tower} fixes the spacing of the descendant tower to $2/\ap$.

\section{Conclusion and Outlook}
\label{sec:outlook}

We have given a direct route from the full-time representation of a fixed-graph in-in correlator to its tropical and stringy integrals.  For a connected loopless multigraph $G$, the piecewise-linear action $\calS_G^V$ is the support function of the in-in zonotope $\Zono(G)$.  Replacing its site and edge terms by the corresponding positive Laurent binomials gives $\calI_G^{\zono}(\ap)$.  Thus the zonotope and its finite-$\ap$ deformation are not separate constructions: both follow from the same time integral, according to whether the Laurent factors are tropicalized or retained.

The vertex-space integral gives $\corrG_V=\Omega_G^{\zono}$, the stripped graph-dependent correlator.  The remaining propagator factors $1/(2y_e)$ arise from elementary edge-space beta integrals, so the complete stringy correlator is $\calI_G^{\square}\calI_G^{\zono}$.  In the field-theory limit the associated geometry is therefore $\Zono(G)$ times a centered hypercube.  The zonotope contains all dependence on the adjacency of $G$; the hypercube keeps track of the normalization of the individual propagators.

The augmented graph $\widehat G{=}G\vee\{0\}$ gives a uniform rational form for this vertex-space integral.  A simplicial ordering-cone refinement of the normal fan yields the chamber sum for $\Omega_G^{\zono}$.  Individual chamber terms may contain cut energies $\widehat X_A$ associated with disconnected vertex subsets.  These auxiliary denominators cancel in the sum, leaving precisely the partial-energy poles $X_S$ of connected vertex-induced subgraphs $G[S]$.

Much of this graph structure persists at finite $\ap$.  Parallel edges combine through the sum of their energies, giving the exact melonic reduction~\eqref{eq:melonic-reduction}.  Setting edge energies to zero gives the deletion identity~\eqref{eq:edge-deletion}, with a product over connected components when the deletion disconnects the graph.  For the field-theory correlator, the corresponding edge-pole residue also carries the factor $2^{-|I|}$ supplied by the explicit propagators.  Near a partial-energy pole $X_S=0$, the stringy integral separates into an internal block for $G[S]$ and shifted zonotope integrals for the components of its complement.  The internal block reduces to beta functions when $G[S]$ is a tree, while cycles leave a coupled integral; the same expansion produces the descendant tower $X_S=-2m/\ap$.

Appendix~\ref{app:tree-cubeahedron} records a complementary Born-rule geometry for trees by identifying the graph correlahedron with the graph cubeahedron of $L(G)$, together with its Minkowski-sum realization and stringy integral. Extending this alternative construction to cycles and parallel edges requires additional care and is not needed for the zonotopal result.

The Laurent master function also gives the logarithmic critical equations recorded in Appendix~\ref{app:critical-points}.  After passing to the sign-invariant variables $a_v=z_v^2$, the same divisor defines an affine arrangement with hyperplanes $a_v=0$, $a_v=-1$, and $a_u+a_v=0$. For generic independent logarithmic weights, the finite-field description gives its generic reference degree, together with closed expressions for paths, stars, and cycles. These degrees are arrangement-theoretic reference quantities and are not assumed to coincide with the critical-point count after the physical exponent specialization.

The results above suggest several directions for extending the fixed-graph framework.  A first step is to pass from a single graph to the sum over all graphs contributing in a specified theory, for example conformally coupled $\operatorname{Tr}(\phi^3)$ theory, and to ask whether this sum admits a single tropical integral together with a positive Laurent lift.  It would be interesting to understand whether the resulting Newton geometry is related to, or provides a stringy realization or refinement of, the cosmohedron or correlator-polytope constructions \cite{Arkani-Hamed:2024jbp,Ardila-Mantilla:2026cbo,Figueiredo:2025daa}. Already at fixed graph, there is a complementary question concerning the two geometric descriptions encountered in this paper.  The full-time representation leads directly to the vertex-space in-in zonotope, whereas the Born rule leads to the edge-space $\mathcal G$-cubeahedron .  Since these constructions encode the same fixed-graph correlator in rather different variables and admit different finite-$\ap$ lifts, it would be useful to understand whether they are related more directly, for example through a subdivision, projection, or pushforward of their canonical forms \cite{Glew:2026von,Figueiredo:2025daa}.

At finite $\ap$, several questions remain even within the zonotopal construction.  Beyond the locations of the descendant poles $X_S=-2m/\ap$, it would be interesting to determine their residues and whether they obey a closed factorization or recursion generalizing the $m=0$ partial-energy residue. More broadly, one can ask which analogues of hidden zeros~\cite{Arkani-Hamed:2023swr,De:2025bmf}, kinematic splits~\cite{Cao:2024gln,Cao:2024qpp}, differential relations between theories~\cite{Cheung:2017ems,Backus:2025njt,Dong:2025lrf}, and other field-theory identities persist in the stringy deformation. One may further ask whether poles, descendant residues, zeros, and suitable asymptotic data are sufficient to characterize $\calI_G^{\zono}(\ap)$ recursively.  Finally, the logarithmic critical equations derived here suggest a possible CHY-type~\cite{Cachazo:2013hca} formulation of cosmological correlators.  This would require understanding the critical-point count under the physical exponent specialization and constructing a pushforward from the physical critical points to $\Omega_G^{\zono}=\corrG_V$, in analogy with the CHY representation of scattering amplitudes and the critical-point formulation of stringy canonical forms~\cite{ArkaniHamedHeLamThomas2019}.

\acknowledgments
We would like to thank Dongyu Yang for collaboration at an early stage of this project.  The work of S.H. is supported by the National Natural Science Foundation of China under Grant Nos.~12225510 and 12247103, and by the New Cornerstone Science Foundation.  The work of F.Z. is supported in part by the Science Challenge Project (No.~TZ2025012) and NSAF No.~U2330401.

\appendix

\section{The \texorpdfstring{$\mathcal G$}{G}-cubeahedral alternative for tree graphs}
\label{app:tree-cubeahedron}
We now describe an alternative stringy representation of the fixed-graph correlator for a tree graph.  Since deleting $k$ edges from a tree $G$ produces $k+1$ connected components, every term in the Born-rule expansion~\eqref{eq:born-rule-correlator} carries the same overall weight of $2$:
\begin{equation}
    \corrG^{\rm tree}=2\sum_{I\subseteq E(G)} \frac{\prod_{K\in \pi_0(G\setminus I)}\Psi_{K}} {\prod_{e\in I}y_e}\,.
    \label{eq:tree-born-rule-correlator}
\end{equation}
Each term also contains the common factor $2\prod_{v\in V(G)}p_v^{-1}$.  After stripping this factor, the remaining terms are encoded by the \emph{graph correlahedron}~\cite{Figueiredo:2025daa}. Its facets are the edge poles $y_e$ and the partial-energy poles $X_I$, with $I\in\calB(G)$, where $\calB(G)$ denotes the collection of nonempty connected edge sets of $G$.  The $X_I$ facets meet according to the compatibility of the tubes $G[I]$.  The remaining compatibility rules are
\begin{equation}
	y_e\sim y_f\quad(e\neq f), \qquad y_e\sim X_I\quad\Longleftrightarrow\quad e\notin I.
	\label{eq:tree-graph-correlahedron-compatibility}
\end{equation}

\begin{figure}[htbp]
	\centering
	\includegraphics[width=0.85\linewidth]{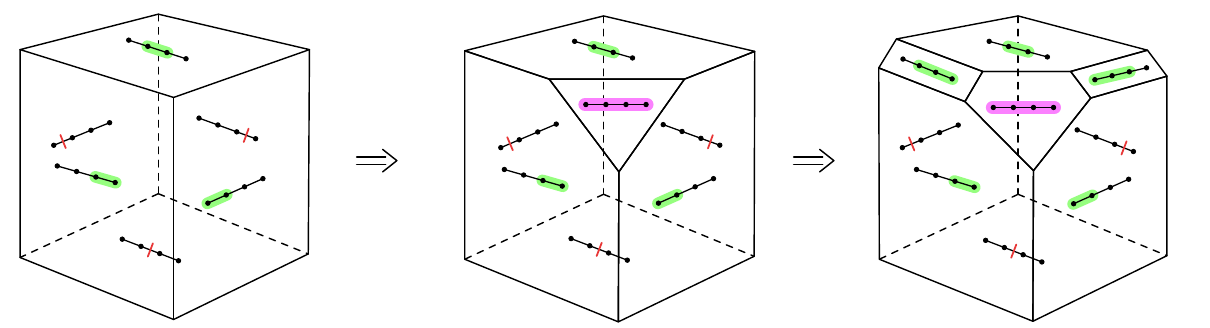}
		\caption{The truncation construction of $\Cube(P_4)$. The initial cube carries square tubes (red bars) and singleton round tubes (green).  The first cut exposes the total round tube (magenta); the remaining cuts produce the other round-tube facets and turn the total-energy facet into $\Asso(P_4)$.  The round tube labeling each truncating facet is the union of the singleton round tubes carried by the facets of the original cube that meet along the truncated face.}
	\label{fig:cubeahedron-truncation}
\end{figure}

Remarkably, for a tree $G$ these are precisely the compatibility relations for the \emph{graph cubeahedron} of $L(G)$ introduced in~\cite{DevadossHeathVipismakul2011}.  The edge pole $y_e$ corresponds to a square tube, while $X_I$ corresponds to a round tube.  Thus, for any tree graph $G$,
\begin{equation}
	\operatorname{GraphCorr}(G) \simeq \operatorname{Cube}(L(G)) =: \Cube(G).
	\label{eq:tree-g-cubehedron-definition}
\end{equation}
The graph cubeahedron can also be understood from the truncation picture of \cite{DevadossHeathVipismakul2011}.  Begin with $[0,1]^{E(G)}$, assigning the facets adjacent to the origin to the edge poles $y_e$, while the opposite facets carry the singleton round tubes $X_{\{e\}}$.  Cutting off the opposite vertex $(1,\dots,1)$ exposes a simplex.  The subsequent cuts indexed by $I\in\calB(G)$, $|I|\geq2$, turn this simplex into $\Asso(G){=}\operatorname{Asso}(L(G))$ while producing the full cubeahedron.  This distinguished associahedral facet is the total-energy facet $X_G$, and therefore carries the fixed-graph wavefunction geometry. This sequence is shown in Figure~\ref{fig:cubeahedron-truncation} for $G=P_4$.

\begin{figure}[htbp]
	\centering
	\includegraphics[width=1\linewidth]{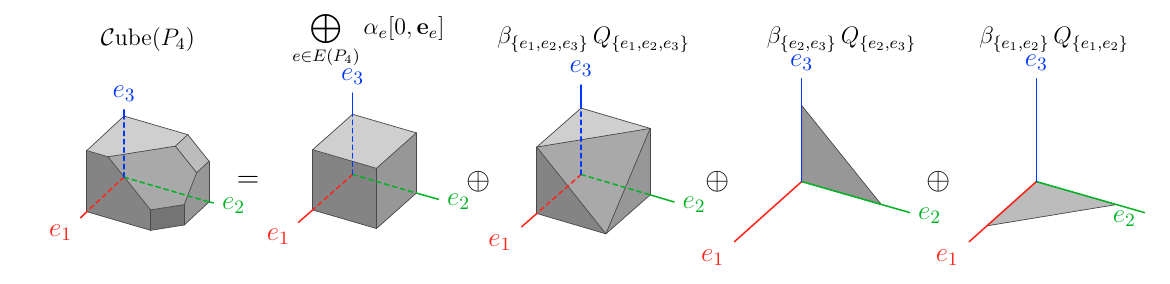}
	\caption{Minkowski decomposition of the $\mathcal G$-cubeahedron $\Cube(P_4)$.}
	\label{fig:cubea_minkowski_P4}
\end{figure}

A Minkowski-sum realization of graph cubeahedra is known \cite{Almeter2022PGraphAssociahedra}.  In the present edge-space coordinates we use
\begin{equation}
	\begin{gathered}
		\Cube(G) = \bigoplus_{e\in E(G)}\alpha_e[0,\bfe_e] \oplus \bigoplus_{\substack{I\in\calB(G)\\ |I|\geq2}}\beta_I Q_I, \qquad \alpha_e,\beta_I>0, \\[-2pt]
		Q_I:=\Conv\left\{\sum_{e\in J}\bfe_e:J\subsetneq I\right\}.
	\end{gathered}
	\label{eq:tree-cube-minkowski-sum}
\end{equation}
Here $Q_I$ is the coordinate cube on $I$ with its top vertex cut off. Since every summand contains the origin as a vertex, the origin remains the uncut vertex of the original hypercube.  The corresponding Minkowski decomposition for $G=P_4$ is shown in Figure~\ref{fig:cubea_minkowski_P4}.

For a genuine Minkowski sum, the parameters in~\eqref{eq:tree-cube-minkowski-sum} are positive and independent.  Matching its square and round support numbers to the physical poles instead selects the following, generally signed, point in parameter space.  For $e=(u,v)$,
\begin{equation}
	\begin{aligned}
		\alpha_e &:=\sum_{\substack{r\in\{u,v\}\\ \deg_G(r)=1}}p_r-y_e, \\
		\beta_I &:=
		\begin{cases}
			(-1)^{|I|}p_v, & I\subseteq\Inc_G(v)\text{ for some }v,\quad |I|\geq2,\\
			0,&\text{otherwise}.
		\end{cases}
	\end{aligned}
	\label{eq:tree-cube-physical-coefficients}
\end{equation}
The parameter assignments above follow from M\"obius inversion~\cite{Zhu:2026vod} relating the parameters to the energy variables labeling the facets; we omit the details here.

\begin{figure}[htbp]
	\centering
	\includegraphics[width=0.6\linewidth]{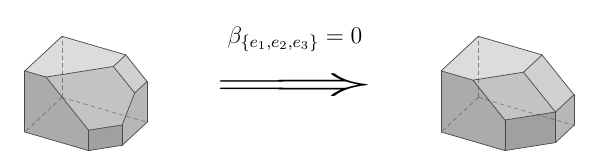}
	\caption{The degeneration of $\Cube(P_4)$ while the weight of Minkowski summand $Q_{\{e_1,e_2,e_3\}}$ becomes zero.}
	\label{fig:cubea_degen}
\end{figure}

The physical specialization may therefore lie on the boundary of the positive parameter space and degenerate the generic cubeahedron.  For $G=P_4$, \eqref{eq:tree-cube-physical-coefficients} gives $\beta_{\{e_1,e_2,e_3\}}=0$.  Equivalently, the weight of the maximal Minkowski summand $Q_{\{e_1,e_2,e_3\}}$ vanishes, producing the degeneration shown in Figure~\ref{fig:cubea_degen}. For this specialization, the degenerate polytope is non-simple, or equivalently, its normal fan contains a non-simplicial cone. This is analogous to passing from the simplicial ordering-cone refinement associated with $\widehat G$ back to the generally non-simplicial normal cones of $\Zono(G)$: in both cases, several simplicial cones merge into a single non-simplicial cone.

We can now uplift the positive family \eqref{eq:tree-cube-minkowski-sum} to a stringy integral.  Introduce positive edge-space coordinates $w_e$ and
\begin{equation}
	F_I(w) :=\prod_{e\in I}(1+w_e)-\prod_{e\in I}w_e =\sum_{J\subsetneq I}\prod_{e\in J}w_e.
	\label{eq:tree-cube-truncated-polynomial}
\end{equation}
Since $\Newt(1+w_e)=[0,\bfe_e]$ and $\Newt(F_I)=Q_I$, \eqref{eq:tree-cube-minkowski-sum} is their weighted Newton polytope.  The corresponding stringy integral~\cite{ArkaniHamedHeLamThomas2019} is
\begin{equation}
	\begin{aligned}
		\calI_G^{\cube}(\ap) &:=\ap^{|E(G)|} \int_{\Rpos^{|E(G)|}} \prod_{e\in E(G)}\dlog w_e\, \prod_{e\in E(G)}w_e^{\ap y_e} \prod_{e\in E(G)}(1+w_e)^{-\ap\alpha_e} \prod_{\substack{I\in\calB(G)\\ |I|\geq2}} F_I(w)^{-\ap\beta_I}.
	\end{aligned}
	\label{eq:tree-cube-stringy-integral}
\end{equation}
The right-hand side is first defined for $\alpha_e,\beta_I>0$ with $(y_e)$ in the interior of \eqref{eq:tree-cube-minkowski-sum}, and is then meromorphically continued to the physical coefficients \eqref{eq:tree-cube-physical-coefficients}.  Its square and round facet variables are $y_e$ and $X_I$, respectively. Restoring the stripped one-site poles and the common factor of two gives the fixed-tree correlator in the field-theory limit,
\begin{equation}
	\corrG =\frac{2}{\prod_{v\in V(G)}p_v} \lim_{\ap\to0^+}\calI_G^{\cube}(\ap).
	\label{eq:tree-cube-correlator-limit}
\end{equation}
In direct analogy with the construction of $\calI_G^{\square}$ in~\eqref{eq:hypercube-stringy-factor}, this prefactor can also be uplifted to an elementary product of beta integrals whose Newton polytope is a vertex-space hypercube. More explicitly, define
\begin{equation}
	\begin{aligned}
		\mathcal H_G^{\rm cube}(\ap) &:=2 \prod_{v\in V(G)} \left[ \ap\int_0^\infty\dlog z_v\, (z_v+z_v^{-1})^{-2\ap p_v} \right]=2 \prod_{v\in V(G)} \frac{\ap}{2} B\left(\ap p_v,\ap p_v\right).
	\end{aligned}
	\label{eq:tree-cube-site-hypercube-factor}
\end{equation}
Then
\begin{equation}
	\lim_{\ap\to0^+}\mathcal H_G^{\rm cube}(\ap) =\frac{2}{\prod_{v\in V(G)}p_v}, \qquad \corrG =\lim_{\ap\to0^+} \mathcal H_G^{\rm cube}(\ap)\calI_G^{\cube}(\ap).
\end{equation}
The Newton polytope of the product is the direct product of the cubeahedron with a centered vertex-space hypercube.  By contrast, the zonotopal Laurent data are written in the physical variables $x_v,y_e$ from the outset and do not require an ambient Minkowski-weight specialization such as \eqref{eq:tree-cube-physical-coefficients}; their energy dependence is still understood meromorphically when residues are taken.

\begin{figure}[htbp]
	\centering
	\includegraphics[width=1\linewidth]{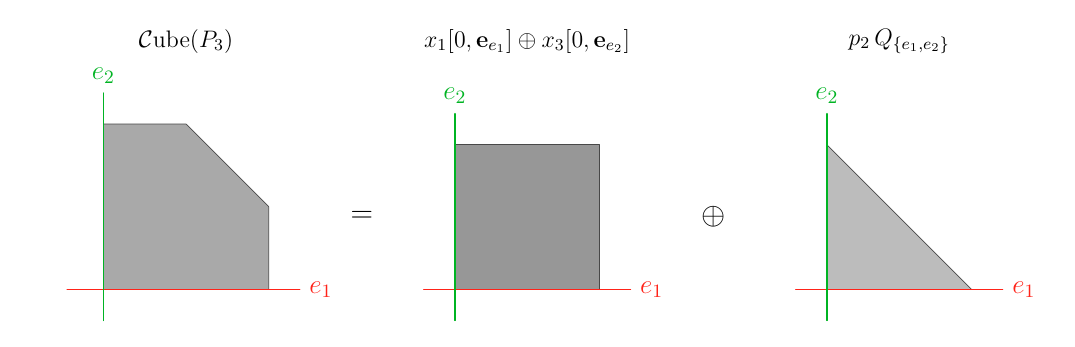}
	\caption{Minkowski decomposition of the $\mathcal G$-cubeahedron $\Cube(P_3)$.}
	\label{fig:cubea_minkowski_P3}
\end{figure}

For a concrete check, take $G=P_3$, with vertices $v_1,v_2,v_3$ and edges $e_1=(v_1,v_2)$, $e_2=(v_2,v_3)$.  Write $x_i:=x_{v_i}$, $p_i:=p_{v_i}$, $y_i:=y_{e_i}$, and $w_i:=w_{e_i}$.  The physical coefficients and the only nontrivial polynomial are
\begin{equation}
	\alpha_{e_1}=x_1, \qquad \alpha_{e_2}=x_3, \qquad \beta_{\{e_1,e_2\}}=p_2, \qquad F_{\{e_1,e_2\}}=1+w_1+w_2.
	\label{eq:p3-cube-physical-data}
\end{equation}
Thus
\begin{equation}
	\begin{aligned}
		\calI_{P_3}^{\cube}(\ap) ={}&\ap^2\int_{\Rpos^2}\dlog w_1\,\dlog w_2\, w_1^{\ap y_1}w_2^{\ap y_2} (1+w_1)^{-\ap x_1}(1+w_2)^{-\ap x_3} (1+w_1+w_2)^{-\ap p_2}.
	\end{aligned}
	\label{eq:p3-cube-stringy-integral}
\end{equation}
Its weighted Newton polygon is (see Figure~\ref{fig:cubea_minkowski_P3})
\begin{equation}
	x_1[0,\bfe_{e_1}] \oplus x_3[0,\bfe_{e_2}] \oplus p_2\Conv(0,\bfe_{e_1},\bfe_{e_2}),
	\label{eq:p3-cube-newton-polygon}
\end{equation}
a pentagon whose five facet variables are
\begin{equation}
	y_1,\quad y_2,\quad X_{\{e_1\}}=x_1+p_2-y_1,\quad X_{\{e_2\}}=x_3+p_2-y_2,\quad X_{\{e_1,e_2\}}=X_G=x_1+x_2+x_3.
	\label{eq:p3-cube-facet-variables}
\end{equation}

The construction above reproduces the tree-level correlator.  For graphs with cycles or parallel edges, however, a naive extension of the same cubeahedral integral does not in general reproduce the cut-dependent powers of two in the Born-rule expansion.  In special cases, such as the $C_2$ graph considered in~\cite[App.~D]{Figueiredo:2025daa}, part of this mismatch can be absorbed into a redefinition of an appropriate tube variable.  This mechanism already fails for the multigraph consisting of two parallel edges between $v_1$ and $v_2$, together with a third edge between $v_2$ and $v_3$: no redefinition of selected tube variables can make the weights consistent across all cuts.  It remains open whether a more general cubeahedral construction can reproduce every Born-rule term for a general graph with parallel edges or cycles.  We make no general no-go claim here; such extensions, together with further properties of cubeahedral stringy forms, are left for separate work.

\clearpage

\section{Critical Equations and Generic Arrangement Degrees}
\label{app:critical-points}
For completeness, we record two related but logically distinct structures associated with the zonotopal stringy integral.  First, its physical master function determines a logarithmic critical-point system with the constrained exponents fixed by the energies $x_v$ and $y_e$.  Second, after complexification and passage to the sign-invariant variables $a_v=z_v^2$, the same divisor defines an affine hyperplane arrangement.  Replacing the physical exponents by generic independent logarithmic weights gives a well-defined generic critical-point degree of this arrangement.

The latter is an arrangement-theoretic reference degree.  We do not assume that it equals the number of complex critical points after imposing the physical exponent specialization, nor do we construct a critical-point pushforward to the correlator.  The generic degree is included here as a mathematical property of the divisor arrangement associated with the stringy integral.

\subsection{Critical equations from the master function}
The zonotopal stringy integral can be written exactly in Euler form as
\begin{equation}
	\calI_G^{\zono}(\ap) = \ap^{|V(G)|} \int_{\Rpos^{V(G)}} \prod_{v\in V(G)}\dlog z_v\, \Phi_G(z)^{\ap},
	\label{eq:zonotope-master-integral}
\end{equation}
where
\begin{equation}
	\Phi_G(z) = \prod_{v\in V(G)}z_v^{p_v} \prod_{v\in V(G)}(1+z_v^2)^{-x_v} \prod_{e=uv\in E(G)}(z_u^2+z_v^2)^{-y_e}.
	\label{eq:zonotope-master-function}
\end{equation}
The overall power $\ap$ multiplying $\log\Phi_G$ does not affect the location of its critical points, so the logarithmic critical equations are
\begin{equation}
	\dd\log\Phi_G(z)=0.
	\label{eq:critical-equations-dlog}
\end{equation}
Although the original integration contour is $\Rpos^{V(G)}$, the algebraic critical-point problem is naturally considered on the complexified complement
\begin{equation}
	M_G^{(z)} := \left\{ z\in(\mathbb C^\ast)^{V(G)}: 1+z_v^2\neq0,\quad z_u^2+z_v^2\neq0 \right\}.
	\label{eq:z-arrangement-complement}
\end{equation}

Equivalently, for every $v\in V(G)$,
\begin{equation}
	0= z_v\frac{\partial}{\partial z_v}\log\Phi_G = p_v -\frac{2x_vz_v^2}{1+z_v^2} -\sum_{e=uv\in\Inc_G(v)} \frac{2y_ez_v^2}{z_v^2+z_u^2},
	\label{eq:logarithmic-master-equations}
\end{equation}
which can also be written as
\begin{equation}
	0= x_v\frac{z_v^2-1}{z_v^2+1} + \sum_{e=uv\in\Inc_G(v)} y_e\frac{z_v^2-z_u^2}{z_v^2+z_u^2}.
	\label{eq:rational-scattering-equations}
\end{equation}
In the full-time variables $z_v=e^{\tau_v}$, the same equations take the hyperbolic form
\begin{equation}
	0=x_v\tanh\tau_v +\sum_{e=uv\in\Inc_G(v)}y_e\tanh(\tau_v-\tau_u), \qquad v\in V(G).
	\label{eq:hyperbolic-scattering-equations}
\end{equation}

Passing to the sign-invariant variables $a_v:=z_v^2$ gives the complex complement
\begin{equation}
	M_G^{(a)} := \left\{ a\in\mathbb C^{V(G)}: a_v\neq0,-1,\quad a_u+a_v\neq0\ \text{for every }uv\in E(G) \right\}.
	\label{eq:a-arrangement-complement}
\end{equation}
The excluded loci $a_v=0$, $a_v=-1$, and $a_u+a_v=0$ form an affine hyperplane arrangement associated with $G$.

Parallel edges do not define distinct hyperplanes in the $a$-variables. Indeed, if a parallel bundle $B$ joins $u$ and $v$, then
\begin{equation}
	\prod_{e\in B}(a_u+a_v)^{-y_e} = (a_u+a_v)^{-y_B}, \qquad y_B:=\sum_{e\in B}y_e.
\end{equation}
Thus the divisor arrangement and its generic degree depend only on the simple skeleton, in agreement with the parallel-edge reduction of Section~\ref{sec:melonic}.  In the remainder of this appendix we therefore combine every parallel bundle into a single edge and, for simplicity, continue to denote the resulting graph by $G$ and its edge energy by $y_{uv}$.

In these variables the physical master function becomes
\begin{equation}
	\Phi_G(a) = \prod_{v\in V(G)}a_v^{p_v/2} \prod_{v\in V(G)}(1+a_v)^{-x_v} \prod_{uv\in E(G)} (a_u+a_v)^{-y_{uv}}.
	\label{eq:a-variable-master-function}
\end{equation}

Hence the logarithmic weights associated with the hyperplanes $a_v=0$, $a_v=-1$, and $a_u+a_v=0$ are $p_v/2,~-x_v,~ -y_{uv}$ respectively. They are not independent because $p_v$ is fixed by \eqref{eq:connected-cut-variable}. The counting problem below instead treats the logarithmic weights of these hyperplanes as independent and generic.

\subsection{Generic arrangement degrees from finite fields}
We now treat the logarithmic weights of the hyperplanes of this affine arrangement as independent and generic.  Let $N_G^{(a)}$ denote the number of isolated complex critical points on $M_G^{(a)}$ for generic weights.  The critical-point--Euler-characteristic relation gives~\cite{OrlikTerao1995,Huh2013MLDegree}
\begin{equation}
	N_G^{(a)} = (-1)^{|V(G)|}\chi\!\left(M_G^{(a)}\right).
	\label{eq:generic-degree-euler}
\end{equation}

The squaring map
\begin{equation}
	\pi:M_G^{(z)}\longrightarrow M_G^{(a)}, \qquad (z_v)_{v\in V(G)} \longmapsto (a_v=z_v^2)_{v\in V(G)},
	\label{eq:z-to-a-cover}
\end{equation}
is an unramified cover of degree $2^{|V(G)|}$.  Hence the corresponding lifted generic count in the complexified $z$-space is
\begin{equation}
	N_G^{(z)} = 2^{|V(G)|}N_G^{(a)}.
	\label{eq:z-a-solution-count}
\end{equation}
This factor is the degree of the algebraic covering and should not be interpreted as a multiplicity of saddles on the original positive-real integration contour.

The Euler characteristic in \eqref{eq:generic-degree-euler} can be obtained from the characteristic polynomial of the arrangement by the finite-field method~\cite{Athanasiadis1996FiniteFields}.  For a sufficiently large odd prime power $q$, define
\begin{equation}
    S_q:=\mathbb F_q\setminus\{0,-1\},
\end{equation}
and let
\begin{equation}
	T_G(q) := \#\left\{ (a_v)_{v\in V(G)}\in S_q^{V(G)}: a_u+a_v\neq0 \quad \text{for every }uv\in E(G) \right\}.
	\label{eq:finite-field-count}
\end{equation}
For all sufficiently large $q$ of good reduction, $T_G(q)$ is the characteristic polynomial of this affine arrangement evaluated at $q$. Therefore
\begin{equation}
	N_G^{(a)} = (-1)^{|V(G)|}T_G(1).
	\label{eq:ml-degree-finite-field}
\end{equation}
Here $T_G(1)$ means that the finite-field count is first identified with its polynomial in $q$ and only then evaluated at $q=1$.

For the graph families below, the finite-field count can be organized by a transfer matrix.  Introduce basis vectors $\lvert a\rangle$, $a\in S_q$, and define
\begin{equation}
	\langle b\rvert A\lvert a\rangle =
	\begin{cases}
		1,&a+b\neq0,\\
		0,&a+b=0.
	\end{cases}
	\label{eq:zonotope-transfer-matrix}
\end{equation}
Then
\begin{equation}
	T_G(q) = \sum_{\{a_v\in S_q\}_{v\in V(G)}} \prod_{uv\in E(G)} \langle a_u\rvert A\lvert a_v\rangle.
	\label{eq:finite-field-transfer-count}
\end{equation}

The value $a=1$ is distinguished because its forbidden partner $-1$ is already absent from $S_q$.

For the three graph families used below, define
\[
\lvert\mathbf 1\rangle := \sum_{a\in S_q}\lvert a\rangle.
\]

\paragraph{Chains.}
For the path $P_n$ on $n$ vertices, the $n-1$ edges give $n-1$ successive transfers, while the values at both endpoints are summed.  Hence
\[
T_{P_n}(q) = \langle\mathbf 1\rvert A^{n-1}\lvert\mathbf 1\rangle.
\]
Evaluating the resulting polynomial at $q=1$ gives
\begin{equation}
	N_{P_n}^{(a)} = \frac{(1+\sqrt2)^n+(1-\sqrt2)^n}{2}.
	\label{eq:path-pell-degree}
\end{equation}
The lifted $z$-system therefore has
\begin{equation}
	N_{P_n}^{(z)} =2^{n-1}\bigl((1+\sqrt2)^n+(1-\sqrt2)^n\bigr).
	\label{eq:path-pell-degree-z}
\end{equation}

\paragraph{Stars.}
Consider the star $K_{1,n}$, with one central vertex and $n$ leaves.  Fix the central value to be $a\in S_q$.  Since there are no edges between the leaves, their values can then be summed independently, giving
\begin{align}
	T_{K_{1,n}}(q)=& \sum_{a\in S_q} \bigl(\langle\mathbf 1\rvert A\lvert a\rangle\bigr)^n = (q-2)^n+(q-3)^{n+1}, \nonumber\\
	& N_{K_{1,n}}^{(a)} = 2^{n+1}-1.
	\label{eq:star-degree}
\end{align}
Since $|V(K_{1,n})|=n+1$, the corresponding count in the $z$-variables is
\begin{equation}
	N_{K_{1,n}}^{(z)} =2^{n+1}\bigl(2^{n+1}-1\bigr).
	\label{eq:star-degree-z}
\end{equation}
The two terms distinguish the central value $a=1$, for which all $q-2$ leaf values are allowed, from the remaining $q-3$ central values, each of which allows $q-3$ leaf values.

\paragraph{Cycles.}
For the simple cycle $C_n$, the value propagated through all $n$ edges must return to its starting value.  Closing the matrix contraction therefore gives
\[
T_{C_n}(q)=\tr(A^n).
\]
Its evaluation at $q=1$ yields
\begin{equation}
	N_{C_n}^{(a)} = (1+\sqrt2)^n+(1-\sqrt2)^n-(-1)^n-2, \qquad n\geq3.
	\label{eq:cycle-degree}
\end{equation}
Thus
\begin{equation}
	N_{C_n}^{(z)} =2^n\left[(1+\sqrt2)^n+(1-\sqrt2)^n-(-1)^n-2\right].
	\label{eq:cycle-degree-z}
\end{equation}

The chain, star, and cycle formulas are thus three implementations of the same finite-field contraction: an open matrix product for $P_n$, independent branch contractions for $K_{1,n}$, and a closed trace for $C_n$.

\clearpage
\bibliographystyle{JHEP}
\bibliography{references}

\end{document}